\documentclass[aps,twocolumn,preprintnumbers,amsmath,amssymb,nofootinbib,superscriptaddress,notitlepage]{revtex4-1}

\usepackage{color}
\usepackage[utf8]{inputenc}
\usepackage{epsfig}
\usepackage{slashed}

 \newcommand{\mpv}[1]{{#1}}

\begin{document}

\title{Peaks within peaks and the possible two-peak structure of
  the $P_c(4457)$: \\ the effective field theory perspective}

\author{Fang-Zheng Peng}

\author{Jun-Xu Lu}
\affiliation{School of Physics, Beihang University, Beijing 100191, China}

\author{Mario S\'anchez S\'anchez}
\affiliation{Centre d'\'Etudes Nucl\'eaires, CNRS/IN2P3, Universit\'e de Bordeaux, 33175 Gradignan, France} 
 
\author{Mao-Jun Yan}
\affiliation{School of Physics,
Beihang University, Beijing 100191, China} 

\author{Manuel Pavon Valderrama}\email{mpavon@buaa.edu.cn}
\affiliation{School of Physics,
  Beihang University, Beijing 100191, China}
\affiliation{International Research Center for Nuclei and Particles
  in the Cosmos and \\
  Beijing Key Laboratory of Advanced Nuclear Materials and Physics, \\
  Beihang University, Beijing 100191, China} 

\date{\today}


\begin{abstract} 
  \rule{0ex}{3ex}
  The LHCb pentaquarks --- the $P_c(4312)$, $P_c(4440)$ and $P_c(4457)$ ---
  have been theorized to be $\Sigma_c \bar{D}$ and
  $\Sigma_c \bar{D}^*$ S-wave molecules.
  Here we explore the possibility that two of these pentaquarks ---
  the $P_c(4440)$ and $P_c(4457)$ --- contain in addition
  a $\Lambda_c(2595) \bar{D}$ component in P-wave. 
  We will analyze the effects of this extra channel within
  two effective field theories:
  the first one will be a standard contact-range effective field theory and
  the second one will include the non-diagonal pion dynamics
  connecting the $\Sigma_c \bar{D}^*$ and $\Lambda_c(2595) \bar{D}$ channels,
  which happens to be unusually long-ranged.
  The impact of the coupled-channel dynamics between the $\Sigma_c \bar{D}^*$
  and $\Lambda_c(2595) \bar{D}$ components is modest at best
  for the $P_c(4440)$ and $P_c(4457)$, which will remain
  to be predominantly $\Sigma_c \bar{D}^*$ molecules.
  However, if the quantum numbers of the $P_c(4457)$ are $J^P = \frac{1}{2}^-$,
  the coupled-channel dynamics is likely to induce the binding of a
  $\Lambda_c(2595) \bar{D}$ S-wave molecule (coupled to $\Sigma_c \bar{D}^*$
  in P-wave) with $J^P = \frac{1}{2}^+$ and a mass
  similar to the $P_c(4457)$.
  If this is the case, the $P_c(4457)$ could actually be a double peak
  containing two different pentaquark states.
\end{abstract}

\maketitle

\section{Introduction}

The discovery of three pentaquark peaks --- the $P_c(4312)$, $P_c(4440)$
and $P_c(4457)$ --- by the LHCb collaboration~\cite{Aaij:2019vzc}
raises the question of what is their nature.
A commonly invoked explanation is that they are $\Sigma_c \bar{D}$ and
$\Sigma_c \bar{D}^*$ bound states~\cite{Chen:2019bip,Chen:2019asm,Liu:2018zzu,Liu:2019tjn,Xiao:2019aya,Valderrama:2019chc,Burns:2019iih,Du:2019pij},
which comes naturally from the closeness of the pentaquark peaks to
the corresponding baryon-meson thresholds and also
from the existence of theoretical predictions predating
their observation~\cite{Wu:2010jy,Wu:2010vk,Wu:2010rv,Xiao:2013yca,Wang:2011rga,Yang:2011wz,Karliner:2015ina}.
Yet the evidence that they are molecular is mostly circumstantial
at the moment and other explanations
might very well be possible~\cite{Eides:2019tgv,Wang:2019got,Yang:2020twg,Ferretti:2020ewe,Stancu:2020paw}.

In this manuscript we will explore a modified molecular interpretation of
the $P_c(4440)$ and $P_c(4457)$ pentaquarks and the consequences it entails.
Of course the fundamental idea will still be that these two pentaquarks are
hadronic bound states, but besides the standard $\Sigma_c \bar{D}^*$
interpretation we will also consider the existence of
a $\Lambda_c(2595) \bar{D}$ ($\Lambda_{c1} \bar{D}$ from now on) component
for the $P_c(4440)$ and $P_c(4457)$.
In the isospin-symmetric limit the $\Sigma_c \bar{D}^*$ and
$\Lambda_{c1} \bar{D}$ threshold are located at $4462.2$
and $4459.5\,{\rm MeV}$, respectively, very close to
the masses of the $P_c(4440)$ and $P_c(4457)$.
Thus it is natural to wonder whether the $\Lambda_{c1} \bar{D}$
channel plays a role in the description of the pentaquarks.

This idea was originally proposed by Burns~\cite{Burns:2019iih},
who conjectured that the $\Lambda_{c1} \bar{D}$ component
might be important for the binding of
molecular pentaquarks.
Later it was realized that the pion-exchange dynamics mediating
the $\bar{D}^* \Sigma_c \to \bar{D} \Lambda_{c1}$ transition
is unusually long-ranged and in practice takes the form of
a $1/r^2$ potential~\cite{Geng:2017hxc}.
This is indeed a really interesting potential in the sense that it can display
discrete scale invariance when attractive enough~\cite{Bawin:2003dm,Braaten:2004pg,Hammer:2005sa},
which in turn opens the possibility of the existence of hadronic molecules
for which there is a geometric spectrum reminiscent of
the Efimov effect in the three-boson system~\cite{Efimov:1970zz}.
For the hidden charm pentaquarks the strength of the $1/r^2$ potential is
probably not enough to trigger a geometric molecular
spectrum~\cite{Geng:2017hxc},
yet this might very well happen in other two-hadron molecular systems.
Recently, Burns and Swanson have considered
the $\bar{D}^* \Sigma_c \to \bar{D}\Lambda_{c1}$ pion-exchange dynamics
beyond its long-distance behavior, leading to the conclusion that
the $P_c(4457)$ might not be a $\frac{1}{2}^-$
$\bar{D}^* \Sigma_c$ S-wave molecular state
but a $\frac{1}{2}^+$ $\bar{D} \Lambda_{c1}$ one instead~\cite{Burns:2019iih}.

The present manuscript delves further into the consequences that a
$\bar{D} \Lambda_{c1}$ component will have
for the pentaquark spectrum.
For this we formulate two effective field theories (EFTs):
a pionless EFT and a {\it half-pionful} EFT.
By half-pionful we denote an EFT which includes
the unusually long-ranged pion dynamics of the
$\bar{D}^* \Sigma_c \to \bar{D} \Lambda_{c1}$ transition,
for which the characteristic length scale is
between $10$ and $20\,{\rm fm}$,
but does not include the pion dynamics of the $\bar{D}^* \Sigma_c$ system,
which has a range in between $1$ and $2\,{\rm fm}$.
We find that the addition of the $\bar{D} \Lambda_{c1}$ channel is
inconsequential if the quantum numbers of the $P_c(4440)$ and
$P_c(4457)$ molecular pentaquarks are
$\tfrac{1}{2}^-$ and $\tfrac{3}{2}^-$,
respectively.
However, if the quantum numbers of the $P_c(4457)$ pentaquark are
$\tfrac{1}{2}^-$ instead, then the existence of a partner state
with a similar mass and quantum numbers
$\tfrac{1}{2}^+$ is very likely.
That is, the $P_c(4457)$ might be a double peak, as happened
with the original $P_c(4450)$ pentaquark discovered in 2015~\cite{Aaij:2015tga}.

The manuscript is structured as follows. In Sect.~\ref{sec:pionless}
we explain how to describe the $\bar{D} \Sigma_c$, $\bar{D} \Sigma_c^*$
and $\bar{D} \Lambda_{c1}$ interactions within a pionless contact-range EFT.
In Sect.~\ref{sec:half-pionful} we introduce the half-pionful theory,
in which we include the pion exchange transition potential
in the $\bar{D} \Sigma_c^*$-$\bar{D} \Lambda_{c1}$
channel.
In Sect.~\ref{sec:Pc-trio} we revisit the description of the LHCb pentaquark
trio within the previous two EFTs.
Finally in Sect.~\ref{sec:conclusions} we present our conclusions.

\section{Pionless Theory}
\label{sec:pionless}

In this section we will derive the lowest-order (LO) contact-range interaction
for the $\bar{D}^* \Sigma_c$-$\bar{D} \Lambda_{c1}$ system.
\mpv{By {\it pionless theory} we specifically refer to an EFT in which pions are
  subleading, instead of an EFT without pions: while the latter is the usual
  meaning of pionless in the nuclear sector (see, e.g.,
  Refs.~\cite{vanKolck:1998bw,Chen:1999tn}),
  in the hadronic sector we are often constrained to LO calculations.
  Thus it makes sense to use the word pionless to describe the LO EFT only,
  which is the part of the theory that we will be using.}
For formulating the LO contact-range Lagrangian
we will find convenient to use the {\it light-quark notation} explained
in detail in Ref.~\cite{Valderrama:2019sid},
but which has been previously used in the literature,
e.g. in Refs.~\cite{Manohar:1992nd,Karliner:2015ina}.
In contrast with the standard superfield notation (see for instance
Ref.~\cite{Falk:1992cx} for a clear exposition) in which we combine
heavy hadrons with the same light-quark spin into a unique superfield,
in the light-quark notation we simply write the interactions
in terms of the light-quark spin degrees of freedom
within the heavy hadrons.
Of course both notations are equivalent, but for non-relativistic problems
the light-quark notation is easier to use.

\subsection{The $\bar{D} \Sigma_c$ and $\bar{D}^* \Sigma_c$ channels}

The $\bar{D}$ and $\bar{D}^*$ charmed antimesons are $\bar{Q} q$ states
where the light-quark $q$ and heavy antiquark $\bar{Q}$ are in S-wave.
From heavy-quark spin symmetry (HQSS) we expect the heavy-antiquark to
effectively behave as a static color source, which in practical terms
means that the wave function of the light quark is independent of
the total spin of the S-wave heavy meson.
That is, the light-quark wave function (the ``brown muck'') of the $\bar{D}$
and $\bar{D}^*$ charmed antimesons is the same (modulo corrections
coming from the heavy-antiquark mass $m_Q$,
which scale as $\Lambda_{\rm QCD} / m_Q$, with $\Lambda_{QCD} \sim (200-300)\,{\rm MeV}$ the QCD scale).
Two possible formalisms to express this symmetry are the standard heavy-superfield notation and the light-subfield notation.
In the former, we combine the $\bar{D}$ and $\bar{D}^*$ field
into a single superfield~\cite{Falk:1992cx}
\begin{eqnarray}
  H_{\bar{Q}} &=& \frac{1}{\sqrt{2}}\,\left[ \bar{D} +
    \vec{\sigma} \cdot \vec{\bar{D}}^* \right] \, ,
\end{eqnarray}
where the superfield is well-behaved with respect to heavy-antiquark rotations
\begin{eqnarray}
  H_{\bar{Q}} &\to& e^{i \vec{S}_H \cdot \vec{\theta}} H_{\bar{Q}} \, ,
\end{eqnarray}
with $\vec{S}_H$ representing the heavy-antiquark spin operator and
$\vec{\theta}$ the rotation axis and angle.
Thus the combination of $H_{\bar Q}^{\dagger}$ and $H_{\bar Q}$ superfields
in the Lagrangian effectively results in invariance with respect to
heavy-antiquark rotations, i.e. to heavy-antiquark spin.

Conversely, in the light-subfield (or light-quark) notation, we prescind of
writing down the heavy antiquark explicitly and instead
express everything in terms of the effective light-quark degrees of freedom within
the charmed antimeson and the light quark spin operator:
\begin{eqnarray}
  \bar{D} \, , \,  \bar{D}^*  \quad \to \quad q_L  \, , \, \vec{\sigma}_L \, ,
\end{eqnarray}
where $q_L$ represents an effective light-quark subfield,
i.e. a field with the quantum numbers
of the light quark within the charmed antimeson~\footnote{The $q_L$ field 
  does not represent an actual light-quark field $q$,
  but the effective field that results from ignoring
  the heavy-quark spin degree of freedoms
  within the charmed antimeson.}.
Then we write down explicit rules for transforming the light-quark spin
operator into charmed antimeson spin operators
\begin{eqnarray}
  \langle \bar{D} | \vec{\sigma}_L | \bar{D} \rangle &=& 0 \, , \\
  \langle \bar{D} | \vec{\sigma}_L | \bar{D}^* \rangle &=& \vec{\epsilon}_1 \, ,
  \\
  \langle \bar{D}^* | \vec{\sigma}_L | \bar{D}^* \rangle &=& \vec{S}_1 \, ,
\end{eqnarray}
where $\vec{\epsilon}_1$ is the polarization vector of
the $\bar{D}^*$ meson and $\vec{S}_1$ the spin-1 matrices.

Regarding the $\Sigma_c$ and $\Sigma_c^*$ charmed baryons,
their quark content is $Q q q$ where the $q q$ diquark
has light spin $S_L = 1$ and the system is in S-wave.
The structure of the S-wave charmed baryons is independent of whether the
baryon spin is $S = \tfrac{1}{2}$ ($\Sigma_c$) or
$\tfrac{3}{2}$ ($\Sigma_c^*$).
In the standard heavy-superfield notation this is taken into account
by defining the superfield~\cite{Cho:1992cf}
\begin{eqnarray}
  \vec{S}_Q &=& \frac{1}{\sqrt{3}}\,\vec{\sigma}\,\Sigma_c + \vec{\Sigma}_c^*
  \, ,
\end{eqnarray}
which has good heavy-quark rotation properties, while in the light-quark
notation we simply write everything in terms of the light-diquark
subfield $a_L$ (i.e. the $qq$ pair) and
its light-spin operator
\begin{eqnarray}
  \Sigma_c \, , \,  \Sigma_c^*  \quad \to \quad a_L  \, , \, \vec{S}_L \, ,
\end{eqnarray}
where $a_L$ is the field representing the effective light-diquark
degrees of freedom (with quantum numbers $J^P = 1^+$,
i.e. an axial vector), with the translation rules
\begin{eqnarray}
  \langle \Sigma_c | \vec{S}_L | \Sigma_c \rangle &=&
  \frac{2}{3}\,\vec{\sigma}_2  \, , \label{eq:trans-baryon-1} \\
  \langle \Sigma_c | \vec{S}_L | \Sigma_c^* \rangle &=& \frac{1}{\sqrt{3}}
  \vec{S}_2 \, , \label{eq:trans-baryon-2}
  \\
  \langle \Sigma_c^* | \vec{S}_L | \Sigma_c^* \rangle &=&
  \frac{2}{3}\vec{\Sigma}_2 \, , \label{eq:trans-baryon-3}
\end{eqnarray}
where $\vec{\sigma}_2$ are the Pauli matrices as applied to the charmed baryon
$\Sigma_c$, $\vec{S}$ a set the matrices representing the spin-$\tfrac{1}{2}$
to spin-$\tfrac{3}{2}$ transition (which can be consulted
in Ref.~\cite{Lu:2017dvm}) and $\vec{\Sigma}_2$
the spin-$\tfrac{3}{2}$ matrices.

With these ingredients
the interaction between a $\bar{D}$ charmed antimeson and a
$\Sigma_c$ charmed baryon can be easily written as
\begin{eqnarray}
  \mathcal{L}_1 = C_a\,(q_L^{\dagger} q_L)\,(a_L^{\dagger} a_L) 
  +
  C_b\,(q_L^{\dagger} \vec{\sigma}_{L} q_L) \cdot (a_L^{\dagger} \vec{S}_{L} a_L)
  \, , \nonumber \\
\end{eqnarray}
which leads to the non-relativistic contact-range potential
\begin{eqnarray}
  V_{C1} = C_a + C_b\,\vec{\sigma}_{L1} \cdot \vec{S}_{L2} \, .
  \label{eq:V_C1}
\end{eqnarray}
This potential can be particularized for the two cases of interest for us
in the present work, the $\bar{D} \Sigma_c$ and $\bar{D}^* \Sigma_c$
systems
\begin{eqnarray}
  V_{C1}(\bar{D} \Sigma_c) &=& C_a \, , \\
  V_{C1}(\bar{D}^* \Sigma_c) &=& C_a +
  C_b\,\frac{2}{3}\,\vec{S}_1 \cdot \vec{\sigma}_2  \, ,
\end{eqnarray}
which we will use for the $P_c(4312)$ and the $\bar{D}^* \Sigma_c$ component
of the $P_c(4440)$ and $P_c(4457)$, respectively.

\subsection{The $\bar{D}^* \Sigma_c$-$\bar{D} \Lambda_{c1}$ transition}

Now we will consider the $\bar{D}^{(*)} \Sigma_c^{(*)}$ to
$\bar{D}^{(*)} \Lambda_{c1}^{(*)}$ transitions,
which are necessary for the description of the $\bar{D} \Lambda_{c1}$ component
in the $P_c(4440)$ and $P_c(4457)$ pentaquarks.
First we will consider the structure of the $\Lambda_{c1}$ and
$\Lambda_{c1}^*$ P-wave charmed baryons, which are $Q qq$ states
in which the spin of the light-quark pair is $S_L = 0$ and
their orbital angular momentum is $L_L = 1$, yielding
a total angular momentum of $J_L = 1$.
In practice this means that there is no substantial difference
(except for parity) between the description of
the $\Sigma_c$, $\Sigma_c^*$ and $\Lambda_{c1}$, $\Lambda_{c1}^*$ charmed baryons
either in terms of heavy-superfield or light-subfield notations.
In the superfield notation, we will write~\cite{Cho:1994vg}
\begin{eqnarray}
  \vec{R}_Q &=& \frac{1}{\sqrt{3}}\,\vec{\sigma}\,\Lambda_{c1} +
  \vec{\Lambda}_{c1}^* \, ,
\end{eqnarray}
while in the light-quark notation we use
\begin{eqnarray}
  \Lambda_{c1} \, , \,  \Lambda_{c1}^*  \quad \to \quad
  v_L  \, , \, \vec{L}_L \, ,
\end{eqnarray}
with $v_L$ representing the light-diquark pair (with quantum numbers
$J^P = 1^-$, i.e. a vector field) and $\vec{L}_L$ the spin-1 matrices,
where we use a different notation than in the S-wave charmed-baryon case
to indicate that the angular momentum comes
from the orbital angular momentum of
the light-quark pair.
This does not entail any operational difference,
with the translation rules being
\begin{eqnarray}
  \langle \Lambda_{c1} | \vec{L}_L | \Lambda_{c1} \rangle &=&
  \frac{2}{3}\,\vec{\sigma}_2  \, , \label{eq:trans-P-baryon-1} \\
  \langle \Lambda_{c1} | \vec{L}_L | \Lambda_{c1}^* \rangle &=& \frac{1}{\sqrt{3}}
  \vec{S}_2 \, , \label{eq:trans-P-baryon-2}
  \\
  \langle \Lambda_{c1}^* | \vec{L}_L | \Lambda_{c1}^* \rangle &=&
  \frac{2}{3}\vec{\Sigma}_2 \, , \label{eq:trans-P-baryon-3}
\end{eqnarray}
which are analogous to these of the $\Sigma_c$, $\Sigma_c^*$ baryons,
see Eqs.~(\ref{eq:trans-baryon-1}--\ref{eq:trans-baryon-3}).

With these ingredients we are ready to write
the $\bar{D}^{(*)} \Sigma_c^{(*)} \to \bar{D}^{(*)} \Lambda_{c1}^{(*)}$
transition Lagrangian.
If for simplicity we limit ourselves to the subset of operators generating a
$D \to D^*$ transition, we find that at lowest order there are
two independent relevant operators which for convenience we write as
\begin{eqnarray}
  \mathcal{L}_2 &=& D_a\, 
  (q_L^{\dagger} \vec{\sigma}_L  q_L) \, \cdot
  (v_L^{\dagger} \overleftrightarrow{\nabla} a_L) \,
  \nonumber \\ &+&
  i\,D_b\, (q_L^{\dagger} \vec{\sigma}_{L} q_L) \, \cdot \,
  (v_L^{\dagger} \vec{J}_{L} \times  \overleftrightarrow{\nabla}
  a_L) + C.C. \, , \label{eq:SP}
\end{eqnarray}
with $\overleftrightarrow{\nabla} = (\overrightarrow{\nabla} - \overleftarrow{\nabla})$ and
where $\vec{J}_L$ refers to the spin-1 matrices as applied between the
light-diquark axial and vector fields within
the S- and P-wave charmed baryons.
The translation rules for the $\vec{J}_L$ operator happen to be
\begin{eqnarray}
  \langle \Sigma_c | \vec{J}_L | \Lambda_c \rangle &=&
  \frac{2}{3}\,\vec{\sigma}_2  \, , \label{eq:trans-SP-baryon-1} \\
  \langle \Sigma_c | \vec{J}_L | \Lambda_c^* \rangle &=& \frac{1}{\sqrt{3}}
  \vec{S}_2 \, , \label{eq:trans-SP-baryon-2}
  \\
  \langle \Sigma_c^* | \vec{J}_L | \Lambda_c^* \rangle &=&
  \frac{2}{3}\vec{\Sigma}_2 \, , \label{eq:trans-SP-baryon-3}
\end{eqnarray}
which are analogous to
Eqs.~(\ref{eq:trans-baryon-1}--\ref{eq:trans-baryon-3}) and
(\ref{eq:trans-P-baryon-1}--\ref{eq:trans-P-baryon-3}),
except that now the initial and final baryon states
are different (either the S- to P-wave baryon
transition or vice versa).
Other operators choices are possible in the Lagrangian of Eq.~(\ref{eq:SP}),
but the present one is particularly useful because
the $D_a$ term is pion-like, while the $D_b$ term is $\rho$-like:
they are similar to what we could get from the exchange of
a pion and a $\rho$ respectively,
as we explain in the Appendix~\ref{app:saturation}.
The potential we obtain is
\begin{eqnarray}
  V_{C2}(1 \to 2) = + D_a \, \vec{\sigma}_{L1} \cdot \vec{q} + i\,D_b \,
  \vec{q} \cdot \left( \vec{\sigma}_{L1} \times \vec{J}_{L2}\right) \, ,
  \nonumber \\ 
  \label{eq:V-C2-full}
\end{eqnarray}
while in the other direction it is
\begin{eqnarray}
  V_{C2}(2 \to 1) = - D_a \, \vec{\sigma}_{L1} \cdot \vec{q} + i\,D_b \,
  \vec{q} \cdot \left( \vec{\sigma}_{L1} \times \vec{J}_{L2}\right) \, ,
  \nonumber \\
\end{eqnarray}
where $D_a$ and $D_b$ are real in the convention
we have used to write the potentials.
It is important to notice that $V_{C2}$ is a non-diagonal potential
and can be redefined by a phase
\begin{eqnarray}
  V_{C2}(1 \to 2) &\to& e^{+i \phi}\,V_{C2}(1 \to 2) \, , \\
  V_{C2}(2 \to 1) &\to& e^{-i \phi}\,V_{C2}(2 \to 1) \, ,
\end{eqnarray}
in which case the potential is still self-adjoint.
In the convention above, the p-space partial wave projection is purely real
while the r-space partial wave projection is purely imaginary.
To avoid the inconveniences originating from this fact, when working
in coordinate space we will automatically add the phase $\phi = \pm\pi$
for the non-diagonal potential to be real.

Phenomenologically we expect the $D_a$ and $D_b$ couplings to represent
the exchange of a pseudoscalar and vector mesons, respectively.
However there is no short-range contribution directly attributable to a
pseudoscalar meson: pion exchange is excessively long-ranged as to be
included in the contact-range potential.
For taking this into account, we will devise a power counting in which
the $D_a$ coupling is a subleading order contribution,
while $D_b$ remains leading.
Thus the effective potential we will use from now on will be
\begin{eqnarray}
  V_{C2} = i\, D_b \,
  \vec{q} \cdot \left( \vec{\sigma}_{L1} \times \vec{S}_{L2}\right) \, .
  \label{eq:V_C3p}
\end{eqnarray}

\subsection{The $\bar{D} \Lambda_{c1}$ channel}

Finally we consider the $\bar{D}^{(*)} \Lambda_{c1}^{(*)}$ system, which enters
the description of the $P_c(4440)$ and $P_c(4457)$ as an additional
(P-wave) component of the wave function.
Yet this meson-baryon system is particularly relevant for a
theoretical pentaquark with quantum numbers $J^P = \tfrac{1}{2}^+$,
for which the most important meson-baryon component
of the wave function will be $\bar{D} \Lambda_{c1}$ in S-wave.

The lowest order interaction
in the $\bar{D}^* \Lambda_{c1}^*$ system happens to be formally identical to the
one for the $\bar{D}^{(*)} \Sigma_{c}^{(*)}$ system, that is
\begin{eqnarray}
  \mathcal{L}_3 = E_a\, (q_L^{\dagger} q_L) \, (v_L^{\dagger} v_L) + 
  E_b\, (q_L^{\dagger} \vec{\sigma}_{L} q_L) \cdot (v_L^{\dagger} \vec{L}_{L} v_L)
  \, , \nonumber \\
\end{eqnarray}
which leads to the potential
\begin{eqnarray}
  V_{C3} = E_a + E_b\,\vec{\sigma}_{L1} \cdot \vec{L}_{L2} \, .
\end{eqnarray}
If we particularize to the $\bar{D} \Lambda_{c1}$ molecule,
we will end up with 
\begin{eqnarray}
  V_{C3}(\bar{D} \Lambda_{c1}) = E_a \, , \label{eq:V_C2p}
\end{eqnarray}
which is a really simple potential, where the coupling $E_a$ is unknown.

\begin{table*}[t]
\begin{tabular}{|c|c|c|c|c|}
\hline\hline
Molecule & Partial Waves & $J^P$ & $\vec{\sigma}_{L1} \cdot \hat{r} $ &
$\hat{r} \cdot (\vec{\sigma}_{L1} \times \vec{S}_{L2})$ 
\\ \hline
$\bar{D} \Lambda_{c1}$ - $\bar{D}^* \Sigma_c$ & $^2S_{{1}/{2}}$-$^2P_{1/2}$-$^4P_{1/2}$ & $\frac{1}{2}^+$ &
$\left(\begin{matrix}
0 & +\frac{1}{\sqrt{3}} & -\sqrt{\frac{2}{3}} \\
+\frac{1}{\sqrt{3}} & 0 & 0 \\ 
-\sqrt{\frac{2}{3}} & 0 & 0
\end{matrix}\right)$
&
$\left(\begin{matrix}
0 & \frac{2 i}{\sqrt{3}} & i\,\sqrt{\frac{2}{3}} \\
-\frac{2 i}{\sqrt{3}} & 0 & 0 \\
-i\,\sqrt{\frac{2}{3}} & 0 & 0 
\end{matrix}\right)$  \\
\hline
\hline
$\bar{D}^* \Sigma_c$ - $\bar{D} \Lambda_{c1}$
& $^2S_{{1}/{2}}$-$^2P_{1/2}$ & $\frac{1}{2}^-$ &
$\left(\begin{matrix}
0 & \frac{1}{\sqrt{3}} \\
\frac{1}{\sqrt{3}} & 0 
\end{matrix}\right)$
&
$\left(\begin{matrix}
0 & -\frac{2 i}{\sqrt{3}} \\
+\frac{2 i}{\sqrt{3}} & 0
\end{matrix}\right)$  \\
\hline
\hline
$\bar{D}^* \Sigma_c$ - $\bar{D} \Lambda_{c1}$
& $^4S_{{3}/{2}}$-$^2P_{3/2}$ & $\frac{3}{2}^-$ &
$\left(\begin{matrix}
0 & \frac{1}{\sqrt{3}} \\
\frac{1}{\sqrt{3}} & 0 
\end{matrix}\right)$
&
$\left(\begin{matrix}
0 & +\frac{i}{\sqrt{3}}  \\
-\frac{i}{\sqrt{3}} & 0 
\end{matrix}\right)$  \\
\hline
\hline\hline
\end{tabular}
\centering \caption{Matrix elements of the vector operators
  for the partial waves we are considering
  in this work.} \label{tab:operators}
\end{table*}

\subsection{Partial-Wave Projection}

For the partial-wave projection of the contact-range potentials (and the one
pion exchange (OPE) potential later on), we will use the spectroscopic
notation $^{2S+1}L_J$ to denote a state with spin $S$, orbital angular
momentum $L$ and total angular momentum $J$.
For the pentaquarks states we are considering --- $P_c$, $P_c\,'$, $P_c^*$ ---
the relevant partial waves are
\begin{eqnarray}
  P_c(\frac{1}{2}^-) &:& {}^2S_{1/2}(\bar{D} \Sigma_c) \, , \label{eq:Pc1-ch} \\
  P_c\,'(\frac{1}{2}^+) &:&
  {}^2S_{1/2}(\bar{D} \Lambda_{c1}) - {}^2P_{1/2}(\bar{D}^* \Sigma_c) -
  {}^4P_{1/2}(\bar{D}^* \Sigma_c) \, , \nonumber \\ \label{eq:Pc-prima-ch} \\
  P_c^*(\frac{1}{2}^-) &:&
  {}^2S_{1/2}(\bar{D}^* \Sigma_c) - {}^2P_{1/2}(\bar{D} \Lambda_{c1}) \, ,
  \label{eq:Pc2-ch} \\
  P_c^*(\frac{3}{2}^-) &:&
  {}^4S_{3/2}(\bar{D}^* \Sigma_c) - {}^2P_{3/2}(\bar{D} \Lambda_{c1}) \, ,
  \label{eq:Pc3-ch}
\end{eqnarray}
where we indicate the relevant meson-baryon channels within parentheses.

\subsection{Momentum-Space Representation}

For the momentum-space representation, we simply project the relevant
contact-range potential into the partial waves of interest.
For the $P_c$ ($\bar{D} \Sigma_c$) pentaquark we simply have
\begin{eqnarray}
  \langle p' | V(P_c) | p \rangle &=& C_a \, . \label{eq:V-Pc1}
\end{eqnarray}
Next, for the two $P_c^*$ configurations ($J=\tfrac{1}{2}$, $\tfrac{3}{2}$)
we have
\begin{eqnarray}
  \langle p' | V(P_c^*, \frac{1}{2}) | p \rangle &=&
  \begin{pmatrix}
    C_a - \frac{4}{3}\,C_b & \frac{2}{\sqrt{3}}\,\frac{2 D_b}{3}\,p \\
    \frac{2}{\sqrt{3}}\,\frac{2 D_b}{3}\, p' & 0
  \end{pmatrix} \, , \label{eq:V-Pc2} \\
  \langle p' | V(P_c^*, \frac{3}{2}) | p \rangle &=&
  \begin{pmatrix}
    C_a + \frac{2}{3}\,C_b & -\frac{1}{\sqrt{3}}\,\frac{2 D_b}{3}\,p \\
    -\frac{1}{\sqrt{3}}\,\frac{2 D_b}{3}\, p' & 0
  \end{pmatrix} \, . \label{eq:V-Pc3}
\end{eqnarray}
Finally for the $P_c'$ ($\bar{D} \Lambda_{c1})$ pentaquark we have
\begin{eqnarray}
  \langle p' | V(P_c') | p \rangle &=&
  \begin{pmatrix}
    E_a & -\frac{2}{\sqrt{3}}\,\frac{2 D_b}{3}\,p &
    -\sqrt{\frac{2}{3}}\,\frac{2 D_b}{3}\,p \\
    -\frac{2}{\sqrt{3}}\,\frac{2 D_b}{3}\,p'
    & 0 & 0 \\
    -\sqrt{\frac{2}{3}}\,\frac{2 D_b}{3}\,p' & 0 & 0
  \end{pmatrix} \, , \nonumber \\ \label{eq:V-Pc-prima}
\end{eqnarray}
which can be simplified to a two-channel form if we take into account
that the two P-wave $\bar{D}^* \Sigma_c$ components can
adopt the configuration
\begin{eqnarray}
  \frac{2}{\sqrt{6}}\,| \bar{D}^* \Sigma_c ({}^2P_{1/2}) \rangle +
  \frac{1}{\sqrt{3}}\,| \bar{D}^* \Sigma_c ({}^4P_{1/2}) \rangle \, ,
\end{eqnarray}
which maximizes the strength of the transition potential and we end up with
\begin{eqnarray}
  \langle p' | V(P_c') | p \rangle &=&
  \begin{pmatrix}
    E_a & -\sqrt{2}\,\frac{2 D_b}{3}\,p \\
    -\sqrt{2}\,\frac{2 D_b}{3}\,p' & 0 
  \end{pmatrix} \, . \nonumber \\
\end{eqnarray}
Notice that this simplification is only possible for the pionless theory at LO:
if we include pion-exchanges or other effects we will have to revert to
the original three-channel representation.

\subsection{Coordinate Space Representation}

We obtain the r-space contact-range potential from Fourier-transforming
the p-space one
\begin{eqnarray}
  V(\vec{r}) = \int \frac{d^3 \vec{q}}{(2 \pi)^3} \, V(\vec{q}) \,
  e^{- i \vec{q} \cdot \vec{r}} \, ,
\end{eqnarray}
which in the case of the $V_{C1}$ and $V_{C3}$ potentials leads to
\begin{eqnarray}
  V_{C1}(\vec{r}) &=& (C_a + C_b\,\vec{\sigma}_{L1} \cdot \vec{S}_{L2})\,
  \delta^{(3)}(\vec{r})\, , \\
  V_{C3}(\vec{r}) &=& (E_a + E_b\,\vec{\sigma}_{L1} \cdot \vec{L}_{L2})\,
  \delta^{(3)}(\vec{r}) \, .
\end{eqnarray}
For the $V_{C2}$ potential, which contains one unit of orbital angular momentum,
the transformation is a bit more involved, resulting in
\begin{eqnarray}
  V_{C2}(1 \to 2) &=& \Big[
    + i D_a \, \vec{\sigma}_{L1} \cdot \vec{\nabla} \nonumber \\
    && - \,D_b \,
  \vec{\nabla} \cdot \left( \vec{\sigma}_{L1} \times \vec{J}_{L2}\right) \Big]\,
  \delta^{(3)}(\vec{r})\, ,
\end{eqnarray}
which can be further simplified by rewriting
\begin{eqnarray}
  \vec{\nabla}\, \delta^{(3)}(\vec{r}) = \hat{r}\,\partial_r \,
  \delta^{(3)}(\vec{r}) \, ,
\end{eqnarray}
leading to
\begin{eqnarray}
  V_{C2}(1 \to 2) &=& \Big[
    + i D_a \, \vec{\sigma}_{L1} \cdot \hat{r} \nonumber \\ &&
    - \,D_b \,
  \hat{r} \cdot \left( \vec{\sigma}_{L1} \times \vec{J}_{L2}\right) \Big]\,
  \partial_r\,\delta^{(3)}(\vec{r})\, .
\end{eqnarray}
This last expression is particularly useful because
the partial wave projection of the $\vec{\sigma}_{L1} \cdot \hat{r}$ and
$\hat{r} \cdot \left( \vec{\sigma}_{L1} \times \vec{J}_{L2}\right)$
is identical to their p-space versions.
Finally we redefine $V_{C2}(1 \to 2)$ by a phase to end up with a purely real
potential:
\begin{eqnarray}
  V_{C2}(1 \to 2) &\to& - i V_{C2}(1 \to 2) \, .
\end{eqnarray}

With the previous conventions and the power counting we use (for which $D_a$ is a subleading order effect), we end up with the r-space potentials
\begin{eqnarray}
  V(\vec{r}; P_c) &=& C_a\,\delta^{(3)}(\vec{r}) \, , \label{eq:Pc1-r} \\
  V(\vec{r}; P_c^*, \frac{1}{2}) &=&
  \begin{pmatrix}
    C_a - \frac{4}{3}\,C_b & \frac{2}{\sqrt{3}}\,\frac{2 D_b}{3}\,\partial_r \\
    \frac{2}{\sqrt{3}}\,\frac{2 D_b}{3}\, \partial_r & 0
  \end{pmatrix} \delta^{(3)}(\vec{r}) \, , \\
  V(\vec{r} ; P_c^*, \frac{3}{2}) &=&
  \begin{pmatrix}
    C_a + \frac{2}{3}\,C_b & -\frac{1}{\sqrt{3}}\,\frac{2 D_b}{3}\,\partial_r \\
    -\frac{1}{\sqrt{3}}\,\frac{2 D_b}{3}\, \partial_r & 0
  \end{pmatrix} \delta^{(3)}(\vec{r}) \, , \nonumber \\ \\
  V(\vec{r}; P_c') &=& 
  \begin{pmatrix}
    E_a & -\sqrt{2}\,\frac{2 D_b}{3}\,\partial_r \\
    -\sqrt{2}\,\frac{2 D_b}{3}\,\partial_r & 0
  \end{pmatrix} \delta^{(3)}(\vec{r}) \, , \nonumber \\ \label{eq:Pc-Lambda-r}
\end{eqnarray}
where for the $P_c'$ pentaquark we have written the simplified
two-channel version of the potential.

\subsection{Regularization and Renormalization}

The contact-range potentials we are using are not well-defined unless
we include a regulator to suppress the unphysical
high-momentum components of the potential.
For the p-space version of the potential this is done with the substitution
\begin{eqnarray}
  \langle p' | V_C | p \rangle \to
  \langle p' | V_{C,\Lambda} | p \rangle \,
  f(\frac{p'}{\Lambda}) \, f(\frac{p}{\Lambda}) \, ,
  \label{eq:reg-gaussian}
\end{eqnarray}
with $f(x)$ a regulator function, for which we will choose a Gaussian,
$f(x) = e^{-x^2}$.
For the r-space version of the potential we will use a delta-shell regulator
\begin{eqnarray}
  \delta^{(3)}(\vec{r}) &\to& \frac{\delta(r - R_c)}{4 \pi R_c^2} \, ,
  \label{eq:reg-delta-shell-s} \\
  \partial_r\,\delta^{(3)}(\vec{r}) &\to&
  \frac{3}{R_c}\,\frac{\delta(r - R_c)}{4 \pi R_c^2} \, ,
  \label{eq:reg-delta-shell-p} 
\end{eqnarray}
with $R_c$ the coordinate space cutoff, where the $3/R_c$ factor
in the derivative of the delta is chosen for its Fourier-transform to be
either $p$ or $p'$ in the $R_c \to 0$ limit after the partial wave projection.

\subsection{Dynamical Equation}
\label{subsec:dynamical}

For finding the location of the bound states we have to iterate
the r- or p-space potentials that we have obtained
within a dynamical equation.
For the r-space potential, we will solve the reduced Schr\"odinger equation
\begin{eqnarray}
  -u_a'' + \sum_b 2 \mu_b V_{ab}(r) u_b(r) && \nonumber \\
  + \frac{L_a (L_a+1)}{r^2}\,u_a(r) &=& - \gamma_a^2 \, u_a(r) \, ,
\end{eqnarray}
where $a$, $b$ are indices we use to represent the different channels
in the molecules we are considering as detailed
in Eqs.~(\ref{eq:Pc1-ch}--\ref{eq:Pc3-ch}),
while $V_{ab}$ is the potential between two channels,
see Eqs.~(\ref{eq:Pc1-r}--\ref{eq:Pc-Lambda-r}),
which is regularized as
in Eqs.~(\ref{eq:reg-delta-shell-s}) and (\ref{eq:reg-delta-shell-p}).
The reduced mass, angular momentum and wave number of a given channel $a$
are represented by $\mu_a$, $L_a$ and $\gamma_a$.
In turn the wave number is given by
$\gamma_a = \sqrt{2 \mu_a (M_{th(a)} - M)}$,
with $M_{th(a)}$ the mass of the two-hadron threshold for channel $a$ and
$M$ the mass of the molecular pentaquark we are predicting.

For the p-space potential we will solve the Lippmann-Schwinger equation
as applied to the pole of the T-matrix, that is:
\begin{eqnarray}
  \phi_a (\vec{p}) = \sum_b \int \frac{d^3 \vec{q}}{(2 \pi)^3}\,
  \frac{\langle \vec{p} \, | V_{ab} | \vec{q} \rangle}
       {M_{th(b)} - M - \frac{q^2}{2 \mu_b}}\,
  \phi_b (\vec{q}) \, , \nonumber \\
\end{eqnarray}
where $a$, $b$ represent the channel, $\phi_a$ is the vertex function
for channel $a$ (where the vertex function is related to
the residue of the T-matrix),
$V_{ab}$ is the potential between two channels,
see Eqs.~(\ref{eq:V-Pc1}--\ref{eq:V-Pc-prima}),
which is regularized according to Eq.~(\ref{eq:reg-gaussian}), and $M$ is the mass of the molecular pentaquark, while $M_{th(a)}$ and $\mu_a$
are the two-hadron threshold and the reduced mass for a given channel $a$.

\section{Half-Pionful Theory}
\label{sec:half-pionful}

The exchange of one pion between the $\bar{D}^* \Sigma_c$
and $\bar{D} \Lambda_{c1}$ channels has the particularity
that its range is extremely enhanced.
The reason is that the pion in the $\Sigma_c \Lambda_{c1} \pi$ and $D^* D \pi$
vertices can be emitted or absorbed almost on the mass-shell, resulting
in an improved range.
Besides, owing to the opposite parity of the $\Sigma_c$ and $\Lambda_{c1}$
baryons, the pion exchange in this vertex is in S-wave.
In combination with the standard P-wave pion in the vertex involving
the charmed mesons, the outcome is that instead of having a central and
tensor forces with orbital angular momentum $L=0$ and $2$ respectively,
we end up with a vector force with $L=1$.
The long-range behavior of the vector force is $1/r^2$,
i.e. an inverse square-law potential, which can trigger a series of
interesting theoretical consequences when the strength of
the potential is above a certain critical value~\cite{Geng:2017hxc}.
Yet, as explained in Ref.~\cite{Geng:2017hxc}, this is probably not the case for the LHCb pentaquarks
as hadronic molecules.

Now, we begin by writing the pion-exchange Lagrangians
for the $\Sigma_c$ to $\Lambda_{c1}$ transition
in the heavy superfield notation:
\begin{eqnarray}
  \mathcal{L}_{HH \pi} &=& \frac{g_1}{\sqrt{2} f_{\pi}}\,
          {\rm Tr}\left[  H^{\dagger}_{\bar Q} \tau_a
            \vec{\sigma} \cdot \vec{\nabla} \pi_a H_{\bar Q}\right]
  \, , \\
  \mathcal{L}_{RS \pi} &=&
  \frac{h_2}{f_{\pi}}\,
  \vec{R}_Q^{\dagger} \,t_a \partial_0 \pi_a \cdot \vec{S}_Q + C.C. \, ,
\end{eqnarray}
which are obtained from the non-relativistic limits of the Lagrangians of Refs.~\cite{Falk:1992cx,Cho:1994vg}.
The light-quark notation version happens to be trivial
\begin{eqnarray}
  \mathcal{L}_{q_L q_L\pi} &=& \frac{g_1}{\sqrt{2} f_{\pi}}\,
  q_L^{\dagger} \,\tau_a \vec{\sigma}_L \cdot \vec{\nabla}\,\pi_a \, q_L\,,
  \\
  \mathcal{L}_{e_L d_L \pi} &=&
  \frac{h_2}{f_{\pi}}\,
  {v}_L^{\dagger} \,t_a \partial_0 \pi_a \, {a}_L + C.C. \, 
\end{eqnarray}

From the previous Lagrangians we can derive the OPE potential in momentum space,
which reads as follows
\begin{eqnarray}
  V_{\rm OPE}(\vec{q}, 1 \to 2) = \frac{g_1 h_2}{\sqrt{2} f_{\pi}^2}
  \,\vec{\tau}_1 \cdot \vec{t}_2\,
  \frac{\omega_{\pi} \, \vec{\sigma}_{L1} \cdot \vec{q}}
       {{\vec{q}\,}^2 + \mu_{\pi}^2}
  \, , \label{eq:OPE}
\end{eqnarray}
where we are indicating that this is the transition potential in the
$\bar{D}^{(*)} \Sigma_c^{(*)} \to \bar{D}^{(*)} \Lambda_{c1}^{(*)}$ direction.
The operator $\vec{\tau}_1 \cdot \vec{t}_2 = \sqrt{3}$
for total isospin $I = \tfrac{1}{2}$ and $0$ otherwise.
The equivalent expression in coordinate space can be obtained
by Fourier-transforming the previous expression,
where in addition we include a phase to follow
the convention of having a purely real transition potential
\begin{eqnarray}
  V_{\rm OPE}(\vec{r}, 1 \to 2) = 
  \vec{\tau}_1 \cdot \vec{t}_2 \, \vec{\sigma}_{L1} \cdot \hat{r} \,
  W_E(r) 
  \, , \label{eq:OPE-decomp}
\end{eqnarray}
with $W_E$ defined as
\begin{eqnarray}
  W_E(r) &=& \frac{g_1 h_2 \omega_{\pi} \mu_{\pi}^2}{4 \pi \sqrt{2} f_{\pi}^2}\,
  \frac{e^{-\mu_{\pi} r}}{\mu_{\pi} r}\,\left( 1 + \frac{1}{\mu_{\pi} r} \right)
  \, . \label{eq:W_E}
\end{eqnarray}
For the couplings we have taken $g_1 = 0.59$ (as deduced from the $D^* \to D \pi$
and $D^* \to D \gamma$ decays~\cite{Ahmed:2001xc,Anastassov:2001cw}), $h_2 = 0.63$ (from the analysis of Ref.~\cite{Cheng:2015naa},
where $h_2$ is extracted from 
$\Gamma(\Lambda_{c1} \to \Sigma_c \pi)$ as measured by CDF~\cite{Aaltonen:2011sf}),
$f_{\pi} = 130\,{\rm MeV}$ and $\omega_{\pi} \simeq (m(\Lambda_{c1}) - m(
\Sigma_1)) \simeq (m(D^*) - m(D)) \simeq m_{\pi}$, with $m_{\pi} = 138\,{\rm MeV}$. Finally $\mu_{\pi} = \sqrt{m_{\pi}^2 - \omega_{\pi}^2} \simeq 0$, a value we will further discuss in the following lines.

\subsection{Infrared regularization}

In the $\mu_{\pi} \to 0$ limit, which is close to the physical situation we are
dealing with and will probably represent a good approximation of it,
the previous OPE potential becomes a $1/r^2$ infinite-range
potential.
In particular the p-space potential reads
\begin{eqnarray}
  V_{\rm OPE}(\vec{q}, 1 \to 2) \to \frac{g_1 h_2}{\sqrt{2} f_{\pi}^2}
  \,\vec{\tau}_1 \cdot \vec{t}_2\,
  \frac{\omega_{\pi} \, \vec{\sigma}_{L1} \cdot \vec{q}}
       {{\vec{q}\,}^2}
  \, ,
\end{eqnarray}
while for the r-space potential we can take this approximation
into account within the function $W_E$
\begin{eqnarray}
  W_E(r) \to \frac{g_1 h_2 \omega_{\pi}}{4 \pi \sqrt{2} f_{\pi}^2}\,
  \frac{1}{r^2}
  \, .
\end{eqnarray}
Of course this is merely an approximation.
What is actually happening is that the modulus of the effective pion mass
$| \mu_{\pi} |$ will be in general considerably smaller than
the pion mass $m_{\pi}$ (or any other hadronic scale
for that matter).
We have $|\mu_{\pi}| \sim (10-20)\,{\rm MeV}$, its concrete value
depending on the specific particle channel under consideration.
In a few particle channels $\mu_{\pi}$ is purely imaginary, indicating
the possibility of decay into the $\bar{D} \Sigma_c \pi$ channel,
and in others it is real.
A detailed treatment of these difference is however outside
the scope of the present manuscript.

Here we will opt for the much easier treatment we were describing above,
that is, to assume that $\mu_{\pi} = 0$.
For taking into account that the range of the OPE potential is actually not
infinite we will include an infrared cutoff.
For the partial-wave projection of OPE in momentum space,
we will introduce an infrared cutoff $\Lambda_{\rm IR}$ in the following way
\begin{eqnarray}
  \langle p' | V_{\rm OPE} | p \rangle \to
  \langle p' | V_{\rm OPE} | p \rangle\,
  \theta (|q_{-}| - \Lambda_{\rm IR})\,
  \theta (|q_{+}| - \Lambda_{\rm IR}) \, ,
  \nonumber \\ \label{eq:p-IR}
\end{eqnarray}
with $q_{-} = p - p'$ and $q_{+} = p + p'$, with the infrared cutoff
chosen within the cutoff window $\Lambda_{\rm IR} = (10-20)\,{\rm MeV}$,
which corresponds with the size of the modulus of
the effective pion mass.
In coordinate space the inclusion of the infrared cutoff $R_{\rm IR}$
will be considerably simpler
\begin{eqnarray}
  V_{\rm OPE}(\vec{r}) \to V_{\rm OPE}(\vec{r})\,\theta (R_{\rm IR} - r) \, , 
\end{eqnarray}
where we will take $R_{\rm IR} = (10-20)\,{\rm fm}$.

Actually the effect of this infrared cutoff is only important if
the strength of the $1/r^2$ potential is equal or larger
to the critical value triggering a geometric spectrum.
This does not happen for any of the pentaquarks we are considering,
at least with the currently known values of the couplings $g_1$ and $h_2$.
However in the $P_c'$ pentaquark the strength is not far away to
that critical value~\cite{Geng:2017hxc}, indicating that in this case the results
will have a larger dependence on the infrared cutoff.

\subsection{Partial-wave projection}

The partial-wave projection of the OPE potential is trivial
for its coordinate space representation:
owing to its clear separation into a radial and angular piece -- Eq.~(\ref{eq:OPE-decomp}) -- it merely requires to consider
the partial wave projection of the vector operator
$\vec{\sigma}_{L1} \cdot \vec{r}$,
which we already showed in Table \ref{tab:operators}.

For the momentum-space representation of the potential
the partial-wave projection is a bit more complex,
yet it can be written as
\begin{eqnarray}
  \langle p'({}^{S'}{L'}_{J'}) | V | p ({}^{S}{L}_J) \rangle &=&
  \frac{g_1 h_2}{\sqrt{2} f_{\pi}^2}
  \,\vec{\tau}_1 \cdot \vec{t}_2\,
  \omega_{\pi} \, \nonumber \\ &\times&
  \langle {}^{S'}{L'}_{J'} | \vec{\sigma}_{L1} \cdot \hat{q} |{}^{S}{L}_J
  \rangle \nonumber \\ &\times&
  \langle p'({L'}) | \, \frac{1}{|\vec{q}\,|}\, | p ({L}) \rangle \, ,
\end{eqnarray}
where the matrix elements of the vector operator are again to be found
in Table \ref{tab:operators}, to which we have to add the partial wave
projection of the $1/|\vec{q}|$ potential:
\begin{eqnarray}
  \langle p'(1) | \, \frac{1}{|\vec{q}\,|}\, | p(0) \rangle = \frac{2 \pi}{p'}\,
  \left[ 1 + \frac{p'^2 - p^2}{2 p p'} \log{\left| \frac{p+p'}{p-p'}
      \right|} \right] \, , \nonumber \\
  \langle p'(0) | \, \frac{1}{|\vec{q}\,|}\, | p(1) \rangle = \frac{2 \pi}{p}\,
  \left[ 1 + \frac{p^2 - p'^2}{2 p p'} \log{\left| \frac{p+p'}{p-p'}
      \right|} \right] \, . \nonumber \\
\end{eqnarray}
Of course, we still supplement the previous expressions
with the infrared cutoff of Eq.~(\ref{eq:p-IR}).

\section{The pentaquark trio revisited}
\label{sec:Pc-trio}

In this Section we consider the description of the $P_c(4312)$, $P_c(4440)$ and
$P_c(4457)$ pentaquarks within the EFTs proposed in this work.
We will begin by reviewing their standard molecular interpretations as
$\bar{D} \Sigma_c$ and $\bar{D} \Sigma_c^*$ bound states and
then we will move to the novel molecular interpretation
in which the $\bar{D} \Lambda_{c1}$ channel is included
as an explicit degree of freedom for the $P_c(4440)$
and $P_c(4457)$ pentaquarks.
The prediction of a $\bar{D} \Lambda_{c1}$ bound state is contingent on
an unknown coupling constant, $E_a$.
For dealing with this issue we will consider two different estimations of
the value of this coupling and the predictions they will entail.

\subsection{The standard molecular interpretation}

We begin by reviewing the standard molecular interpretation
of Ref.~\cite{Liu:2019tjn}, in which the pentaquarks
were considered to be $\bar{D} \Sigma_c$ and $\bar{D}^* \Sigma_c$ molecules
(without any $\bar{D} \Lambda_{c1}$ component) described by a pionless EFT.
This pionless EFT is equivalent to using the $V_{C1}$ contact-range potential of
Eq. (\ref{eq:V_C1}), which contains two independent couplings $C_a$ and $C_b$.
The original procedure~\cite{Liu:2019tjn} for determining
these two couplings was as follows:
\begin{itemize}
\item[(i)] use the $P_c(4440)$ and $P_c(4457)$ as $\bar{D}^* \Sigma_c$ molecules
  to determine the $C_a$ and $C_b$ couplings;
\item[(ii)] postdict the $P_c(4312)$ as a $\bar{D} \Sigma_c$
  molecule and compare with its experimental location.
\end{itemize}
For convenience we will modify the previous procedure in this manuscript:
\begin{itemize}
\item[(i)] use the $P_c(4312)$ and $P_c(4457)$ as $\bar{D} \Sigma_c$ and
  $\bar{D}^* \Sigma_c$ molecules to determine the $C_a$ and $C_b$ couplings;
\item[(ii)] postdict the $P_c(4440)$ as a $\bar{D}^* \Sigma_c$ molecule and
  compare with its experimental location.
\end{itemize}
This choice guarantees that the prediction of the pentaquark trio remain
all below their respective meson-baryon thresholds: the later
inclusion of the $\bar{D} \Lambda_{c1}$ channel
can in a few instances move the $P_c(4457)$
a bit above the threshold for hard cutoffs if we fit the couplings as in Ref.~\cite{Liu:2019tjn}.

Now  for the $\bar{D}^* \Sigma_c$ molecules there are two spin configurations,
$J = \tfrac{1}{2}$ and $\tfrac{3}{2}$, but we do not know which one
corresponds to each of the pentaquarks.
As a consequence we consider two scenarios, $A_0$ and $B_0$:
\begin{itemize}
\item[(a)] in scenario $A_0$ the $P_c(4440)$ has $J = \tfrac{1}{2}$, while
  the spin of the $P_c(4457)$ is $J = \tfrac{3}{2}$,
\item[(b)] in scenario $B_0$ the $P_c(4440)$ has $J = \tfrac{3}{2}$, while
  the spin of the $P_c(4457)$ is $J = \tfrac{1}{2}$,
\end{itemize}
where we use the subscript ``zero'' to indicate that this is the base case
in which the $\bar{D} \Lambda_{c1}$ channel is not included.
Then we postdict the location of the $P_c(4440)$ in each scenario,
resulting in
\begin{eqnarray}
  M_{A_0} &=& 4440.1 \,(4434.5) \, {\rm MeV} \, , \label{eq:p-space-1} \\
  M_{B_0} &=& 4449.6 \,(4447.6)\, {\rm MeV} \, ,  \label{eq:p-space-2}
\end{eqnarray}
for the p-space Gaussian regulator with $\Lambda = 0.5\,(1.0)\,{\rm GeV}$ and
\begin{eqnarray}
  M_{A_0} &=& 4438.9 \,(4433.8)\, {\rm MeV} \, , \label{eq:r-space-1} \\
  M_{B_0} &=& 4449.3 \,(4447.5)\, {\rm MeV} \, , \label{eq:r-space-2}
\end{eqnarray}
for the r-space delta-shell regulator with $R = 1.0\, (0.5)\,{\rm fm}$.
These numbers are to be compared with the experimental value
$M = (4440.3 \pm 1.3 ^{+4.1}_{-4.6}) \,{\rm MeV}$,
which indicates that scenario $A_0$ is preferred
over scenario $B_0$ (particularly for softer cutoffs).
This coincides with the conclusions of the previous
pionless EFT of Ref.~\cite{Liu:2019tjn}.
\mpv{
  Yet it should be noted that this is a ${\rm LO}$ calculation and
  hence we should expect a sizable truncation error, which can be
  estimated by comparing the predictions at different cutoffs,
  for instance.
  The cutoff variation indicates a $\Delta M = 2.0-5.6\,{\rm MeV}$
  uncertainty for the masses of the pentaquarks,
  see Eqs.~(\ref{eq:p-space-1}-\ref{eq:r-space-2}).
  Once we consider this uncertainty, both scenarios $A$ and $B$ happen to
  be compatible with the experimental location of the pentaquarks.
  For comparison purposes, we notice that EFT calculations including
  the $\bar{D}^* \Sigma_c$ pion dynamics~\cite{Valderrama:2019chc,Du:2019pij},
  which we have considered here to be a subleading order effect and
  thus part of $\Delta M$, tend to prefer scenario $B$.
}

\begin{table}[!ttt]
\begin{tabular}{|ccccc|}
\hline \hline
Scenario & $\Lambda$ $({\rm GeV})$ &
$C_a$ $({\rm fm}^2)$ & $C_b$ $({\rm fm}^2)$ & $E_a^{\rm crit}$ $({\rm fm}^2)$ \\
  \hline
  $A_0$ & $0.5$ & $-2.17$ & $+0.55$ & $-1.13$ \\
  $A_0$ & $1.0$ & $-0.80$ & $+0.13$ & $-0.57$ \\
  \hline
  $B_0$ & $0.5$ & $-2.17$ & $-0.27$ & $-1.13$ \\
  $B_0$ & $1.0$ & $-0.80$ & $-0.07$ & $-0.57$ \\
  \hline \hline
  Scenario & $R_c$ $({\rm fm})$  & $C_a$ $({\rm fm}^2)$ & $C_b$ $({\rm fm}^2)$ &
  $E_a^{\rm crit}$ $({\rm fm}^2)$ \\
  \hline
  $A_0$ & $0.5$ & $-0.81$ & $+0.14$ & $-0.58$ \\ 
  $A_0$ & $1.0$ & $-2.16$ & $+0.53$ & $-1.18$\\ 
  \hline
  $B_0$ & $0.5$ & $-0.81$ & $-0.07$ & $-0.58$ \\ 
  $B_0$ & $1.0$ & $-2.16$ & $-0.26$ & $-1.18$ \\
  \hline \hline
\end{tabular}
\caption{
  The contact-range couplings $C_a$, $C_b$ and $E_a$ when the $\bar{D}^* \Sigma_c$ and $\bar{D} \Lambda_{c1}$ channels do not couple.
  $C_a$ and $C_b$ are obtained from the condition of reproducing
  the mass of the $P_c(4312)$ and $P_c(4457)$ as molecular pentaquarks
  in p- and r-space (as indicated by type of cutoff: $\Lambda$ and $R_c$).
  Scenario $A$ (and its variants) corresponds to considering
  that the spin-parities of
  the $P_c(4440)$ and $P_c(4457)$ are $J^P = \tfrac{1}{2}^-$ and
  $\tfrac{3}{2}^-$, respectively, while scenario $B$ corresponds
  to the opposite identification.
  $E_a^{\rm crit}$ is the critical value of the $E_a$ coupling required for the uncoupled $\bar{D} \Lambda_{c1}$ system to bind.
}
\label{tab:couplings-standard}
\end{table}

\subsection{The novel molecular interpretation}

Now we explore the novel molecular interpretation we propose, in which
the $P_c(4312)$ is a $\bar{D} \Sigma_c$ molecule while the $P_c(4440)$
and $P_c(4457)$ are $\bar{D}^* \Sigma_c$-$\bar{D} \Lambda_{c1}$ molecules.
The contact-range piece of the potential for the pionless and
half-pionful EFTs is given by Eqs.~(\ref{eq:V-Pc1}--\ref{eq:V-Pc3}),
which contain three independent coupling constants
($C_a$, $C_b$ and $D_b$).
Finally we conjecture the existence of a $\bar{D} \Lambda_{c1}$ S-wave
molecule, which we call the $P_c'$ and for which the contact-range
piece of the potential is given by Eq.~(\ref{eq:V-Pc-prima}),
which includes a new coupling ($E_a$).

Of these four couplings, we can determine three of them --- $C_a$, $C_b$
and $D_b$ --- from the masses of the three pentaquarks.
The procedure we will follow is:
\begin{itemize}
\item[(i)] use the $P_c(4312)$ as a $\bar{D} \Sigma_c$ molecule to
  determine the $C_a$ coupling,
\item[(ii)] use the $P_c(4440)$ and $P_c(4457)$ as
  $\bar{D}^* \Sigma_c$-$\bar{D} \Lambda_{c1}$ molecules
  to determine the $C_b$ and $D_b$ couplings,

\item[(ii')] if there is no solution for the previous procedure,
  we will set $D_b = 0$ and, as in the uncoupled-channel case,
  we will determine $C_b$ from the condition of reproducing
  the $P_c(4457)$ pole,
  
\item[(iii)] finally we determine for which values of
  $E_a$ the $P_c'$ (the conjectured
  S-wave $\bar{D} \Lambda_{c1}$ molecule) binds and compare these values
  with expectations from naive dimensional analysis (NDA).
\end{itemize}
As in the standard molecular interpretation, we have two possible scenarios
which we now call $A_1$ and $B_1$, where $A_1$ ($B_1$) corresponds to
the $P_c(4457)$ being a $J = \tfrac{3}{2}$ ($\tfrac{1}{2}$) molecule.
We will further subdivide the scenario $A_1$ ($B_1$) into a pionless
and a half-pionful version, which we will denote
$A_1^{\slashed \pi}$ ($B_1^{\slashed \pi}$) and $A_1^{\pi}$ ($B_1^{\pi}$),
respectively.
It happens that the couplings can be compared with NDA, in particular
$D_b$ and $E_a$: the $D_b$ comparison can provide an indirect estimation
of the likelihood of scenarios $A_1$ and $B_1$, while $E_a$ will provide
the binding likelihood of the $P_c'$ pentaquark.

\begin{table}[!ttt]
\begin{tabular}{|cccccc|}
\hline \hline
Scenario & $\Lambda$ $({\rm GeV})$ & $C_a$ $({\rm fm}^2)$ & $C_b$ $({\rm fm}^2)$ & $D_b$ $({\rm fm}^3)$
& $E^{\rm crit}_a$ $({\rm fm}^2)$  \\
  \hline
  $A_1^{\slashed \pi}$ & $0.5$ & $-2.17$ & $+0.55$ & $0$ & $-1.13$ \\ 
  $A_1^{\slashed \pi}$ & $1.0$ & $-0.80$ & $+0.12$ & $0$ & $-0.57$  \\ 
  \hline
  $B_1^{\slashed \pi}$ & $0.5$ & $-2.17$ & $-0.85$ & $0.99$ & {$+0.04$} \\ 
  {$B_1^{\slashed \pi}$} & $1.0$ & $-0.80$ & $-0.15$ &  {$0.13$} & {$-0.40$} \\ 
  \hline \hline
  $A_1^{\pi}$ & $0.5$ & $-2.17$ & $+0.57$ & $0$ & $-1.04(2)$ \\ 
  $A_1^{\pi}$ & $1.0$ & $-0.80$ & $+0.15$ & $0$ & $-0.50(1)$ \\ 
  \hline
  $B_1^{\pi}$ & $0.5$ & $-2.17$ & $-0.76$ & $1.01$ & $+0.18(1)$\\ 
  $B_1^{\pi}$ & $1.0$ & $-0.80$ & $-0.11$ & $0.12$ & $-0.35$ \\ 
  \hline \hline
  Scenario & $R_c$ $({\rm fm})$ & $C_a$ $({\rm fm}^2)$ &
  $C_b$ $({\rm fm}^2)$ & $D_b$ $({\rm fm}^3)$ & $E^{\rm crit}_a$ $({\rm fm}^2)$ \\
  \hline
  $A_1^{\slashed \pi}$ & $0.5$ & $-0.81$ & $+0.12$ & $0$ & $-0.51$ \\ 
  $A_1^{\slashed \pi}$ & $1.0$ & $-2.16$ & $+0.51$ & $0$ & $-1.06$ \\
  \hline
  $B_1^{\slashed \pi}$ & $0.5$ & $-0.81$ & $-0.14$ & $0.09$ & $-0.42$ \\ 
  $B_1^{\slashed \pi}$ & $1.0$ & $-2.16$ & $-0.75$ & $0.66$ & $-0.15$ \\ 
  \hline \hline
  $A_1^{\pi}$ & $0.5$ & $-0.81$ & $0.15$ &  $0$ & $-0.52(2)$ \\ 
  $A_1^{\pi}$ & $1.0$ & $-2.16$ & $0.55$ &  $0$ & $-1.10(8)$ \\  
  \hline
  $B_1^{\pi}$  & $0.5$ & $-0.81$ & $-0.17$ & $0.09$ & $-0.38(2)$ \\ 
  $B_1^{\pi}$  & $1.0$ & $-2.16$ & $-0.68$ & $0.67$ & $-0.09(8)$ \\ 
  \hline \hline
\end{tabular}
\caption{
  The contact-range couplings $C_a$, $C_b$ and $D_b$ from the condition of
  reproducing the mass of the $P_c(4312)$ and $P_c(4457)$ as molecular
  pentaquarks in p- and r-space (as indicated by type of cutoff:
  $\Lambda$ and $R_c$).
  Scenario $A$ (and its variants) corresponds to considering
  that the spin-parities of the $P_c(4440)$ and $P_c(4457)$
  are $J^P = \tfrac{1}{2}^-$ and $\tfrac{3}{2}^-$, respectively,
  while scenario $B$ corresponds to the opposite identification.
  For the half-pionful EFT, i.e. scenarios $A_1^{\pi}$ and $B_1^{\pi}$,
  the number displayed corresponds to the softer infrared cutoff,
  i.e. $\Lambda_{IR} = 10\,{\rm MeV}$ and $R_{IR} = 20\,{\rm fm},$
  while the number in parenthesis represents the difference
  with $\Lambda_{IR} = 20\,{\rm MeV}$ and $R_{IR} = 10\,{\rm fm}$
  in the last significant digit (if any).
  $E_a^{\rm crit}$ is the critical value of the $E_a$ coupling required
  for the coupled $\bar{D} \Lambda_{c1}$ system to bind.
}
\label{tab:couplings-novel}
\end{table}

To illustrate this idea, we can consider the pionless p-space calculation,
which for $\Lambda = 0.5\,{\rm GeV}$ in scenario $A_1^{\slashed \pi}$ and
$B_1^{\slashed \pi}$ gives
\begin{eqnarray}
  C_a &=& -2.17\,{\rm fm}^2 \, , \\
  \nonumber \\
  C_b &=& +0.55\,{\rm fm}^2 \, \, \mbox{$(A_1^{\slashed \pi})$} \, , \\
  D_b &=& +0.00\,{\rm fm}^2 \, \, \mbox{$(A_1^{\slashed \pi})$} \, ,  \\
  \nonumber \\
  C_b &=& -0.85\,{\rm fm}^2 \, \, \mbox{$(B_1^{\slashed \pi})$} \, , \\
  D_b &=& +0.99\,{\rm fm}^2 \, \, \mbox{$(B_1^{\slashed \pi})$} \, .
\end{eqnarray}
This translates into the following condition for the $P_c'$ to bind
\begin{eqnarray}
  E_a &\leqslant& -1.13\,{\rm fm}^2 \, \, \mbox{$(A_1^{\slashed \pi})$} \, , \\
  E_a &\leqslant& +0.04\,{\rm fm}^2 \, \, \mbox{$(B_1^{\slashed \pi})$} \, ,
\end{eqnarray}
which in scenario $A_1^{\slashed \pi}$ requires the coupling $E_a$
to be attractive, while scenario $B_1^{\slashed \pi}$ will
lead to binding even for a slightly repulsive coupling.
For the calculation we have used the following values for the masses of the hadrons involved: $m(D) = 1867.22\,{\rm MeV}$, $m(D^*) = 2008.61\,{\rm MeV}$, $m(\Sigma_c) = 2453.54\,{\rm MeV}$, $m(\Lambda_{c1}) = 2592.25\,{\rm MeV}$, which are the isospin averages of the PDG values~\cite{Tanabashi:2018oca}.

A complete list of the couplings can be consulted
in Table~\ref{tab:couplings-novel} for the different EFTs
and regulators considered in this work.
Independently of the choice of regulator and cutoff, the binding of the $P_c'$
pentaquark is much more probable in scenario $B_1$ (pionless or half-pionful).
In scenario $A_1$ there is no pair of values for the $C_b$ and $D_b$ couplings that simultaneously reproduces the $P_c(4440)$ and $P_c(4457)$ pentaquarks, and thus we have set $D_b = 0$ and followed the same procedure as in scenario $A_0$ to determine $C_b$.
We will further comment on why this happens later on in this section.

Now we can compare the previous numbers with the NDA estimation of
the expected size of a contact-range coupling
\begin{eqnarray}
  \mpv{|C_{l, l'}| \sim | c_{l,l'} | \, \frac{4 \pi}{M^{l + l' + 2}}} \, ,
  \label{eq:C-NDA}
\end{eqnarray}
where $l$, $l'$ are the angular momenta of the initial and final states
that the contact-range potential couples, \mpv{$M$ is
the hard-scale of the theory and $c_{l,l'} \sim \mathcal{O}(1)$ a numerical
factor of order one related to the partial wave projection.
The origin of this estimation deserves further discussion.
On the one hand we have the $1/M^{l+l'+2}$ scaling, which comes
from the canonical dimension of the contact-range
coupling, i.e. $[C_{l,l'}] = [{\rm mass}]^{-l-l'-2}$.
On the other the numerical factors can be deduced
from different heuristical arguments.
Here we will consider a simple argument based on matching
the $C_{l,l'}$ coupling with the Fourier transform of an unknown,
generic short-range potential characterized by the hard-range scale $M$, i.e.
\begin{eqnarray}
  V_S(r) = M \, g(M r) \, e^{-M r} \, , \label{eq:V_S}
\end{eqnarray}
where we are ignoring further dependence on the angular momentum
for simplicity.
The exponential decay indicates that this potential is generated
by the exchange of a meson of mass $M$ at short distances,
while $g(x)$ a function that depends on the details of
the interaction with that meson.
We expect $C_{l,l'}$ to be proportional to the Fourier transform of $V_S$
at low energies, i.e.
\begin{eqnarray}
  C_{l,l'}\,p^l {p'}^{l'} \propto \lim_{p,p' \to 0}
  \langle p'(l') | V_S | p(l) \rangle \, ,
\end{eqnarray}
with the partial wave projection of the Fourier transform given by
\begin{eqnarray}
  \langle p'(l') | V_S | p(l) \rangle =
  4\pi\,i^{l-l'} \, \int dr\,r^2 j_l(p r) V_S(r) j_{l'}(p' r) \, , \nonumber \\
\end{eqnarray}
with $j_l(x)$ the spherical Bessel function and $V_S$ the short-range potential.
From these two formulas we can easily trace back the $4\pi$ and $c_{l,l'}$
factors in Eq.~(\ref{eq:C-NDA}): the $4\pi$ is simply a trivial
consequence of writing the Fourier transform in the partial
wave basis, while the $c_{l,l'}$ factor comes from the interplay between
the low momentum behavior of the spherical Bessel functions
and the exponential decay of the short-range potential.
For instance, if we consider the short-range S-wave and S-to-P-wave
potentials generated by the exchange of a meson, the general form of
the short-range potential $V_S$ will probably be a Yukawa
for the S-wave case and a derivative of a Yukawa
for the S-to-P wave transition (in the line of Eq.~(\ref{eq:W_E})).
If this is the case, the $g$ functions in Eq.~(\ref{eq:V_S}) will take the form
\begin{eqnarray}
  g_{\rm S}(x) \propto \frac{1}{x} \quad \mbox{and} \quad
  g_{\rm SP}(x) \propto \frac{1}{x}\,(1+\frac{1}{x}) \, ,
\end{eqnarray}
which leads to the numerical factors $c_{S} \propto 1$ and $c_{SP} \propto 1$.
Other arguments might lead to different estimations of the $c_{l,l'}$ factors,
but we will expect most of them to converge
towards $c_{l,l'} = \mathcal{O}(1)$.
For a recent review about NDA in EFTs we recommend Ref.~\cite{vanKolck:2020plz},
which further points out to $c_{l,l'} = \mathcal{O}(1)$.
But we observe that most arguments are devoted to S-wave contact-range
couplings, with comparatively less effort invested in the naturalness
of the $l,l' \neq 0$ cases.
The exception is Halo EFT~\cite{Bertulani:2002sz}, which is however focused
on the more interesting non-natural cases.
The bottom-line is that the NDA estimates for $C_S$ are more
well-established than for $C_{SP}$ and thus conclusions based
on the naturalness of $C_{SP}$ are less robust
than the ones based on $C_S$.
}

\mpv{Here we will choose $c_{l,l'} = 1$ for simplicity.}
For hadrons we expect $M \sim 1\,{\rm GeV}$, which gives us the following
estimations for an S-wave and S-to-P wave counterterms
\begin{eqnarray}
  |C_S^{\rm NDA}| \sim 0.49\,{\rm fm}^2 \quad \mbox{and} \quad
  \mpv{|C_{SP}^{\rm NDA}| \sim 0.10\,{\rm fm}^3} \, .
\end{eqnarray}
From this we see that $C_a$ is unnatural (see Table~\ref{tab:couplings-standard}), which is to be expected
for the coupling of a two-body system that binds~\cite{Birse:1998dk,vanKolck:1998bw},
while $C_b$ and $D_b$ are closer to natural,
though this depends on the cutoff.
\mpv{This is particularly true for $D_b$, see Table~\ref{tab:couplings-novel},
  which is close to its NDA estimate for $\Lambda = 1.0\,{\rm GeV}$
  but not so for $\Lambda = 0.5\,{\rm GeV}$.
  This might be very well related to our choice of
  $M \sim 1\,{\rm GeV}$ for the hard scale, though:
  had we chosen a smaller $M$, we would have ended up with a stronger case
  for the naturalness of $D_b$ (particularly because of the $1/M^3$ scaling).
  Besides this}, we can appreciate that in scenario $A_1$ the binding of
the $P_c'$ pentaquark is possible but not particularly probable,
as the size of the coupling $E_a$ that is required to bind is
larger than the NDA expectation.
In contrast, in scenario $B_1$ the coupling $E_a$ required to bind
falls well within what is expected from NDA.
Thus in this second case binding seems to be much more likely.

\begin{table}[!ttt]
\begin{tabular}{|ccccc|}
\hline \hline
Scenario & $\Lambda$ $({\rm MeV})$ &
$E^{\rm crit}_a$ $({\rm fm}^2)$ & $E^{\rm NDA}_a$ $({\rm fm}^2)$
& $M^{\rm NDA}(P_c')$ \\
  \hline
  $A_1^{\slashed \pi}$ & $0.5$ & $-1.13$ & $-0.49$ & $-$ \\ 
  $A_1^{\slashed \pi}$ & $1.0$ & $-0.57$ & $-0.49$ & $-$ \\
  \hline
  $B_1^{\slashed \pi}$ & $0.5$ & $+0.04$ & $-0.49$ & $4457.0$ \\ 
  $B_1^{\slashed \pi}$ & $1.0$ & $-0.40$ & $-0.49$ & $4457.9$ \\
  \hline \hline
  $A_1^{\pi}$ & $0.5$ & $-1.04(2)$ & $-0.49$ & $-$ \\ 
  $A_1^{\pi}$ & $1.0$ & $-0.50(1)$ & $-0.49$ & $-$ \\
  \hline
  $B_1^{\pi}$ & $0.5$ & $+0.18(1)$ & $-0.49$ & $4456.3$ \\ 
  $B_1^{\pi}$ & $1.0$ & $-0.35$ & $-0.49$ & $4457.0$ \\
  \hline \hline
  Scenario & $R_c$ $({\rm fm})$ &
  $E^{\rm crit}_a$ $({\rm fm}^2)$ & $E^{\rm NDA}_a$ $({\rm fm}^2)$
  & $M^{\rm NDA}(P_c')$ \\
  \hline
  {$A_1^{\slashed \pi}$} & $0.5$ & $-0.58$ & $-0.49$ & $-$ \\ 
  {$A_1^{\slashed \pi}$} & $1.0$ & $-1.18$ & $-0.49$ & $-$ \\
  \hline
  {$B_1^{\slashed \pi}$} & $0.5$ & $-0.42$ & $-0.49$ & $4458.1$ \\ 
  {$B_1^{\slashed \pi}$} & $1.0$ & $-0.15$ & $-0.49$ & $4458.2$ \\
  \hline \hline
  $A_1^{\pi}$ & $0.5$ & $-0.52(2)$ & $-0.49$ & $-$ \\ 
  $A_1^{\pi}$ & $1.0$ & $-1.10(8)$ & $-0.49$ & $-$ \\
  \hline
  $B_1^{\pi}$ & $0.5$ & $-0.38(2)$ & $-0.49$ & $4457.3$ \\ 
  $B_1^{\pi}$ & $1.0$ & $-0.09(8)$ & $-0.49$ & $4457.7$ \\ 
  \hline \hline
\end{tabular}
\caption{
  The mass of the $P_c'$ pentaquark as deduced from the NDA estimate of the
  $E_a$ coupling (assuming it is attractive) in \mpv{scenarios $A_1$ and $B_1$},
  both in the pionless and half-pionful theories.
  For reference, the $\bar{D} \Lambda_{c1}$ threshold is located at
  $4459.5\,{\rm MeV}$ in the isospin-symmetric limit.
  $E_a^{\rm crit}$ has the same meaning as in Table~\cite{tab:couplings-novel}.
}
\label{tab:M-Pc-prima}
\end{table}

Regarding the $P_c'$ pentaquark, we can deduce its probable mass
from the NDA estimation of the $E_a$ coupling,
provided this coupling is attractive:
\begin{eqnarray}
  E_a^{\rm NDA} \simeq -\frac{4 \pi}{M^{2}} \, . \label{eq:Ea-NDA}
\end{eqnarray}
Within scenario $B_1$, this estimation of the coupling consistently generates
a shallow $P_c'$ close to the $\bar{D} \Lambda_{c1}$ threshold,
where the concrete predictions can be consulted
in Table \ref{tab:M-Pc-prima}.
Of course the question is whether it is sensible to assume
that the $E_a$ coupling is attractive.
We will examine the validity of this assumption in the next few lines.

\subsection{Can we further pinpoint the location of the $P_c'$ pentaquark?}

Regarding $E_a$, it will be useful not only to determine its sign
but also its size beyond the NDA estimation we have already
used to argue the existence of the $P_c'$ pentaquark.
From arguments regarding the saturation of contact-range couplings
by light-mesons~\cite{Lu:2017dvm,Peng:2020xrf},
the light-meson contributions to $E_a$ can be divided into two components
\begin{eqnarray}
  E_a = E_a^S + E_a^V \, ,
\end{eqnarray}
which correspond to the scalar ($\sigma$) and vector ($\omega$)
meson contributions.
The scalar and vector contributions are attractive and repulsive
($E_a^S < 0$ and $E_a^V > 0$), respectively.
At first sight this ambiguous result seems to indicate
that we cannot determine the sign of $E_a$,
yet this would be premature.
As a matter of fact the same situation would arise had we applied
this argument to the two-nucleon system, but it happens
that the deuteron binds.
The reason is that the scalar meson contributions have a longer range
than the vector meson ones, leading to net attraction.

This seems to be the case not only in the two-nucleon system,
but also in the $\bar{D} \Lambda_c$ case:
according to a recent calculation in the one-boson-exchange
model~\cite{Chen:2017vai}, 
the $\bar{D} \Lambda_c$ system is not far away from binding.
In fact, had we adapted the recent one-boson exchange model
of Ref.~\cite{Liu:2019zvb} (originally intended for
the $\bar{D}^{(*)} \Sigma_c^{(*)}$ molecules) to
the $\bar{D} \Lambda_{c1}$ system,
the system will not bind, yet its two-body scattering length $a_2$
would probably be unnaturally large
\begin{eqnarray}
  a_2^{\rm OBE}(\bar{D} \Lambda_{c1}) = -24.1^{+20.7}_{-\infty (+9.5)}\,{\rm fm} \, ,
\end{eqnarray}
where the errors are computed as in Ref.~\cite{Liu:2019zvb}
and which are compatible with binding~\footnote{The calculation assumes
  that the $\bar{D}$ and $\Lambda_{c1}$ hadrons are stable, which is not the
  case for the later but neither is this detail important as we take
  the scattering length as a proxy for determining the amount of
  attraction in the two-body system. The cutoff is taken as
  in Ref.~\cite{Liu:2019zvb},
  i.e. $\Lambda = 1.119^{+0.190}_{-0.094}\,{\rm GeV}$,
  while the couplings of the $\Lambda_{c1}$ and $\Sigma_c$ baryons to the
  $\sigma$ and $\omega$ happen to be identical. Finally $\Lambda_{c1}$
  does not couple to the $\rho$, owing to isospin.
}
(the lower error indicates that the scattering
length changes sign, hence the $-\infty$, and that in that case
its value would be $+9.5\,{\rm fm}$).
This reinforces the conclusions derived from Ref.~\cite{Chen:2017vai}
for the $\bar{D} \Lambda_c$ case.
That is, we expect $E_a < 0$ and close to the value required
to have a shallow bound state in the absence of coupling
with the $\bar{D}^* \Sigma_c$ channel.
All this makes the $P_c'$ pentaquark very likely in scenario $B_1$,
as we will now show with explicit calculations.

\begin{figure*}[ttt]
  \begin{center}
    \includegraphics[width=8.2cm]{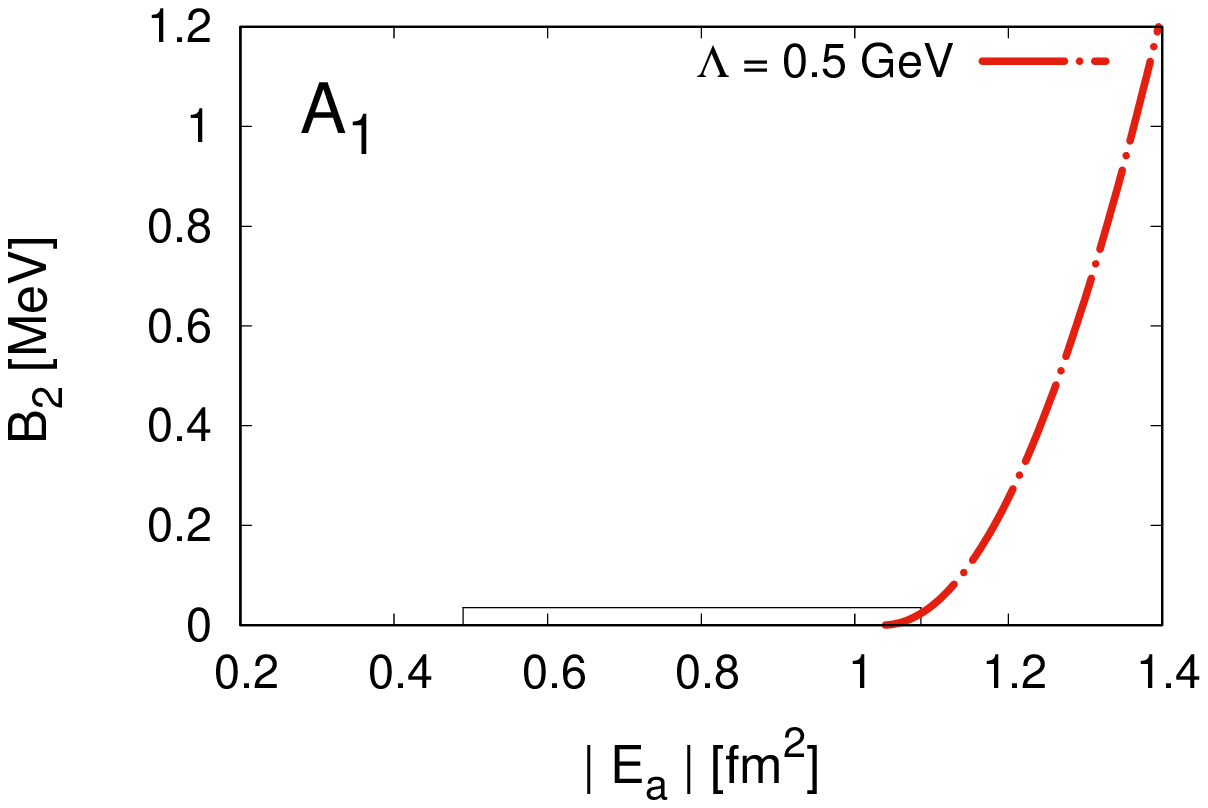}
    \includegraphics[width=8.2cm]{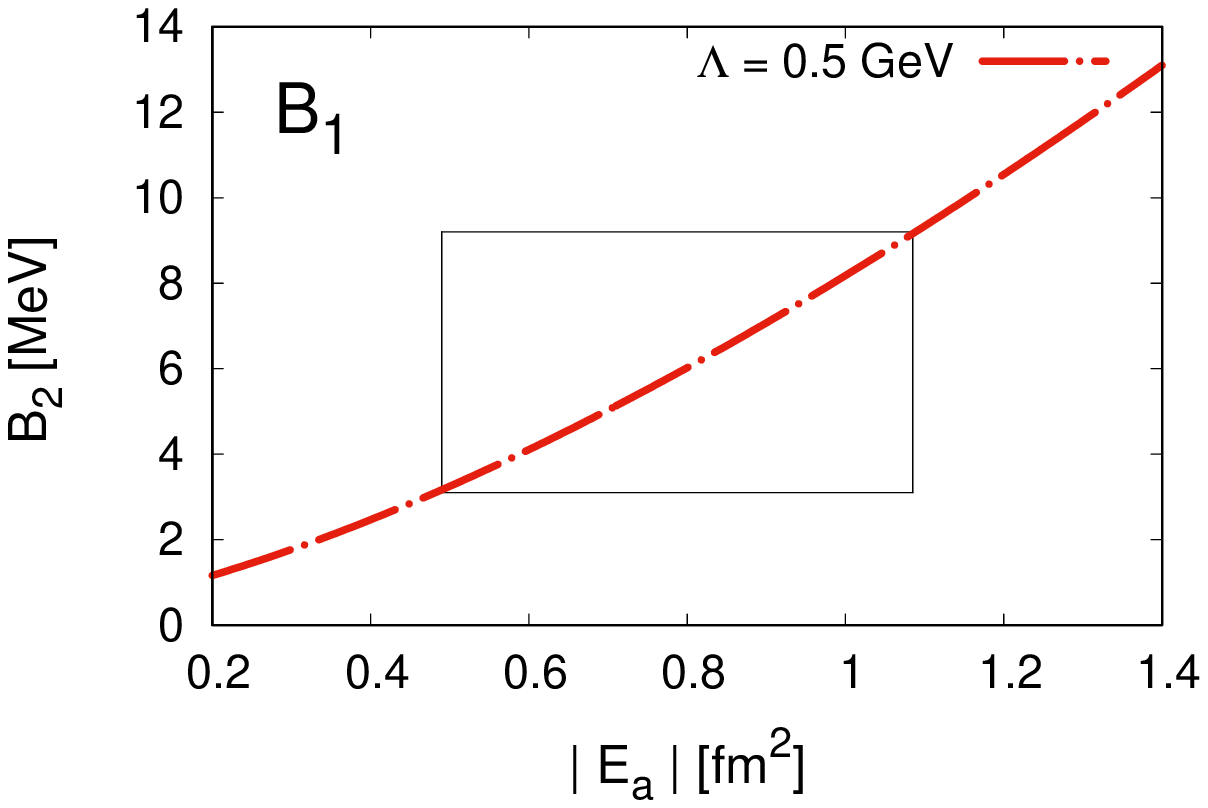}
    \includegraphics[width=8.2cm]{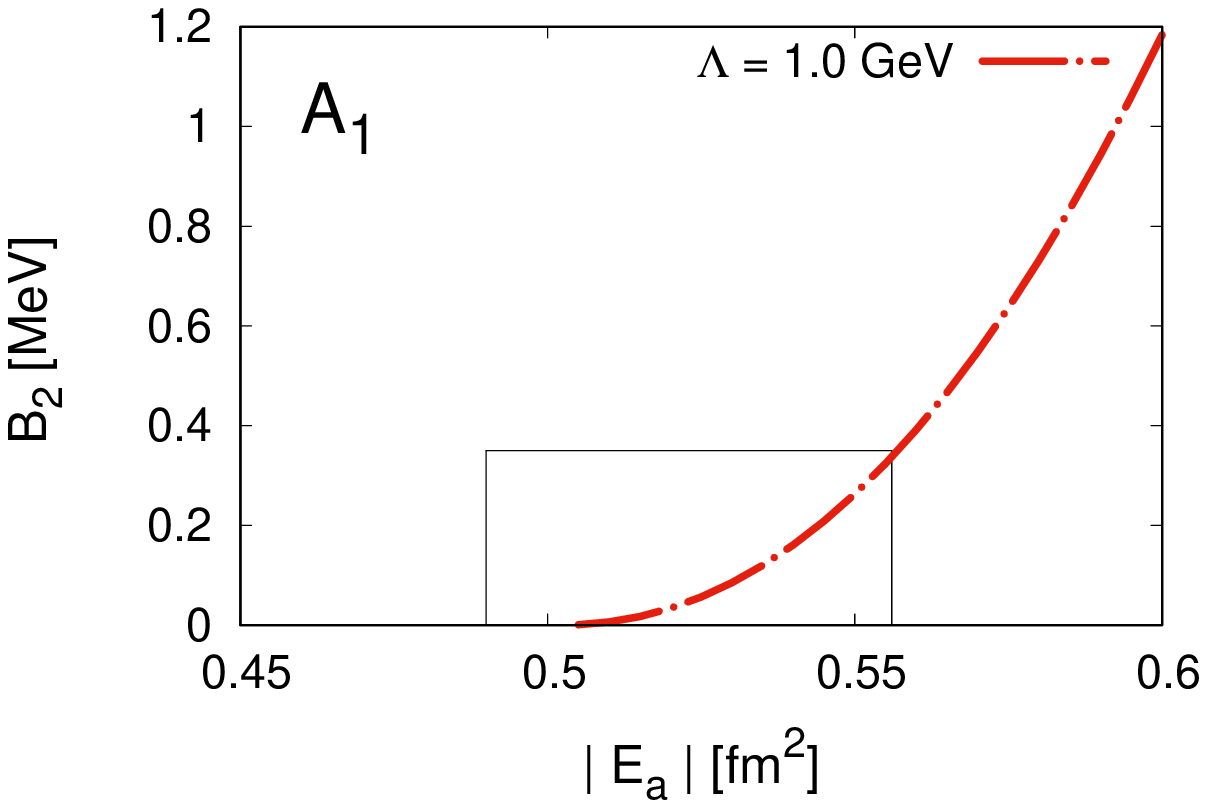}
    \includegraphics[width=8.2cm]{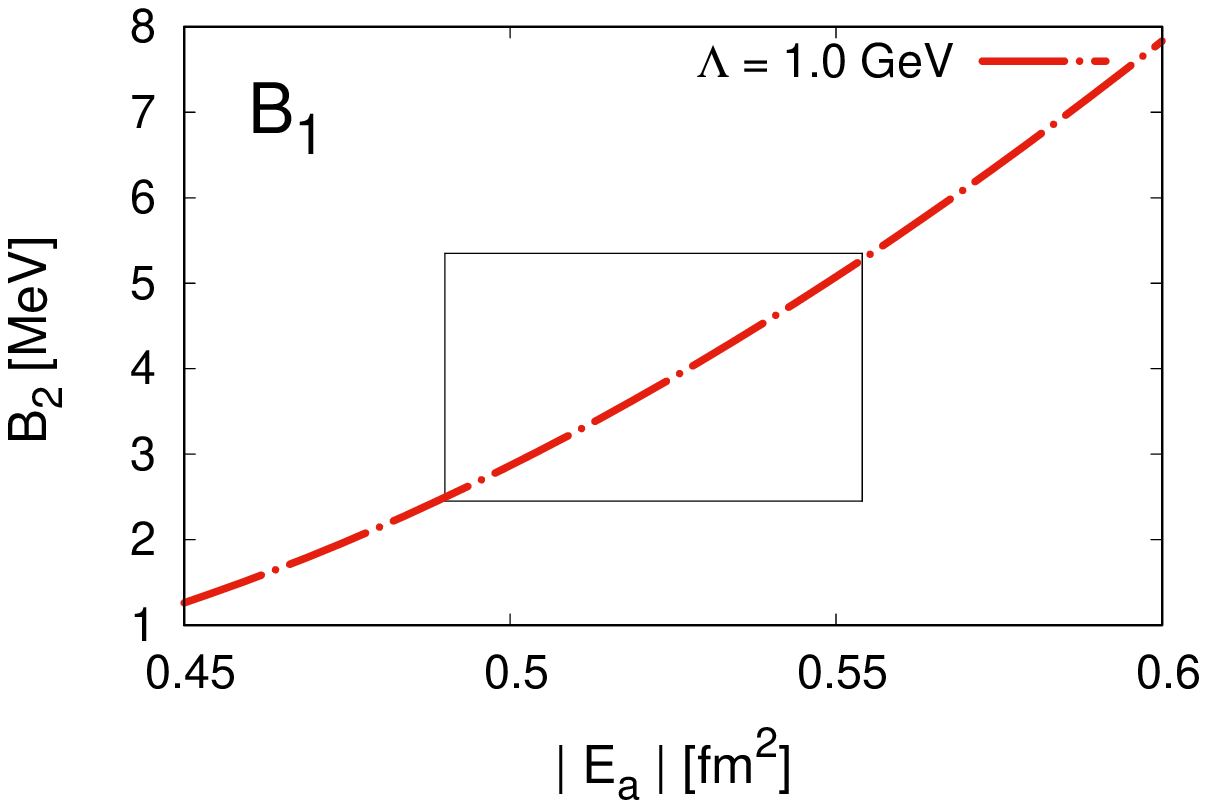}
\end{center}
\caption{
  Binding energy in scenarios $A_1$ and $B_1$ (p-space) of the prospective
  $P_c'$ pentaquark depending on the size of the coupling $E_a$.
  The square indicates what we consider to be the most probable values of
  the coupling $E_a$ and the binding energy of the $P_c'$ pentaquark:
  it comprises the values of $E_a$ from the NDA to the OBE estimations,
  i.e. Eqs.~(\ref{eq:Ea-NDA}) and (\ref{eq:Ea-OBE}).
  For scenario $A_1$ the binding window is rather limited,
  $0-0.03$ ($0-0.3$) ${\rm MeV}$ for $\Lambda = 0.5 (1.0)\,{\rm GeV}$,
  where binding does not happen unless $| E_a |$ is above
  the critical values listed in Table \ref{tab:M-Pc-prima-OBE}.
  For scenario $B_1$ the most probable binding window is about
  $3.2-9.2$ ($2.5-5.3$) ${\rm MeV}$ for $\Lambda = 0.5 (1.0)\,{\rm GeV}$.
  For simplicity we have only shown the half-pionful theory in momentum space
  ($A_1^{\pi}$ and $B_1^{\pi}$), as the other cases considered
  in this work yield similar results (with the curves moving
  a bit towards the right in the pionless case).
}
\label{fig:binding}
\end{figure*}

If we now describe the $\bar{D} \Lambda_{c1}$ two-body system in a pionless EFT,
the coupling $E_a$ can be determined from the value of the scattering length
that we have already computed within the OBE model, leading to
\begin{eqnarray}
  E_a = -1.09^{+0.21}_{-0.18}\,(-0.55^{+0.06}_{-0.04})\,{\rm fm}^2 \, , \label{eq:Ea-OBE}
\end{eqnarray}
for $\Lambda = 0.5\,(1.0)\,{\rm GeV}$ if we do the calculations in p-space,
or alternatively
\begin{eqnarray}
  E_a = -1.10^{+0.21}_{-0.19}\,(-0.56^{+0.06}_{-0.05})\,{\rm fm}^2 \, ,
\end{eqnarray}
for $R_c = 1.0\,(0.5)\,{\rm fm}$ in r-space.
As already explained, this extracted value of the coupling is enough as to
guarantee binding in scenario $B_1$, both in the pionless
and pionful versions.
This would lead to a $P_c'$ that is bound by $(4-9)$ MeV depending
on the case.
The predicted locations can be found in Table \ref{tab:M-Pc-prima-OBE},
where we have \mpv{not} only considered scenario $B_1$ (for which
binding is more probable), \mpv{but also scenario $A_1$
  (for which binding can still happen in the half-pionful case)}.
We can appreciate that \mpv{in scenario $B_1$} the predictions are very similar,
independently of the cutoff or whether the calculation
has been done in r- or p-space.
\mpv{
For a more graphical comparison we have included Fig.~\ref{fig:binding},
which shows the dependence of the binding energy on the coupling $E_a$
for the half-pionful theory in momentum space for scenarios
$A_1$ and $B_1$.
We have chosen this particular calculation as the representative case,
as the other three possible calculations in scenarios $A_1$ and $B_1$
would yield similar results (except that in the pionless theory,
scenario $A_1$ requires a larger $|E_a^{\rm crit}|$ to bind).
In Fig.~\ref{fig:binding} we also indicate the most probable values of $E_a$
and the binding energy of the $P_c'$ within a square.
}

\mpv{At this point it is important to notice that the estimations of
  the $E_a$ coupling discussed here are not only close to binding
  for the $\bar{D} \Lambda_{c1}$ system, but are also compatible
  with it once we take into account the theoretical uncertainties.
  This is not only true for the $E_a$ coupling derived from the OBE model.
  For $\Lambda = 1.0\,{\rm GeV}$ the NDA estimation of $E_a$
  is not far away from the critical value required for binding, i.e.
  $|E_a^{\rm NDA}| = 0.49\,{\rm fm}^2$ to be compared
  with $E_a^{\rm crit} = -0.57\,{\rm fm}^2$.
  The bottom-line is that though the previous discussions have focused
  on scenario $B$, for which binding is more probable,
  the $P_c'$ pentaquark could also exist in scenario $A$.
  That is, the difference between scenarios $A$ and $B$ regarding
  a possible $\bar{D} \Lambda_{c1}$ bound state is merely
  one of likelihood.
}

\begin{table}[!ttt]
\begin{tabular}{|ccccc|}
\hline \hline
Scenario & $\Lambda$ $({\rm GeV})$ &
$E^{\rm crit}_a$ $({\rm fm}^2)$ & $E^{\rm OBE}_a$ $({\rm fm}^2)$
& $M^{\rm OBE}(P_c')$ \\
\hline
  $A_1^{\slashed \pi}$ & $0.5$ & $-1.13$ & $-1.09^{+0.21}_{-0.19}$ & $-$ \\ 
  $A_1^{\slashed \pi}$ & $1.0$ & $-0.57$ & $-0.55^{+0.06}_{-0.04}$ & $-$ \\
\hline
  $B_1^{\slashed \pi}$ & $0.5$ & $+0.04$ &
  $-1.09^{+0.21}_{-0.19}$ & $4451.2^{+2.3}_{-2.2}$ \\ 
  $B_1^{\slashed \pi}$ & $1.0$ & $-0.40$ &
  $-0.55^{+0.06}_{-0.04}$ & $4455.2^{+2.7}_{-2.2}$ \\
\hline \hline
  $A_1^{\pi}$ & $0.5$ & $-1.04(2)$ & $-1.09^{+0.21}_{-0.19}$ & $4459.5^{\dagger}_{-0.6}$ \\ 
  $A_1^{\pi}$ & $1.0$ & $-0.50(1)$ & $-0.55^{+0.06}_{-0.04}$ & $4459.2^{\dagger}_{-0.6}$ \\
\hline
  $B_1^{\pi}$ & $0.5$ & $+0.18$ &
  $-1.09^{+0.21}_{-0.18}$ & $4450.3^{+2.4}_{-2.1}$ \\ 
  $B_1^{\pi}$ & $1.0$ & $-0.35$ &
  $-0.55^{+0.06}_{-0.04}$ & $4454.2^{+2.7}_{-2.3}$ \\
  \hline \hline
  Scenario & $R_c$ $({\rm fm})$ &
  $E^{\rm crit}_a$ $({\rm fm}^2)$ & $E^{\rm OBE}_a$ $({\rm fm}^2)$
  & $M^{\rm OBE}(P_c')$ \\
  \hline
      {$A_1^{\slashed \pi}$} & $0.5$ & $-0.58$ & $-0.56^{+0.06}_{-0.05}$ & $-$ \\ 
      {$A_1^{\slashed \pi}$} & $1.0$ & $-1.18$ & $-1.10^{+0.21}_{-0.19}$ & $-$ \\
    \hline
  $B_1^{\slashed \pi}$ & $0.5$ & $-0.42$ & $-0.56^{+0.06}_{-0.05}$ & $4455.2^{+2.7}_{-2.6}$ \\ 
  $B_1^{\slashed \pi}$ & $1.0$ & $-0.15$ & $-1.10^{+0.21}_{-0.19}$ & $4452.1^{+2.5}_{-2.6}$ \\
    \hline \hline
    $A_1^{\pi}$ & $0.5$ & $-0.52(2)$ & $-0.56^{+0.06}_{-0.05}$ & $4459.2(2)^{\dagger}_{-1.0}$ \\ 
    $A_1^{\pi}$ & $1.0$ & $-1.10(8)$ & $-1.10^{+0.21}_{-0.19}$ & $4459.5^{\dagger}_{-0.6}$ \\
    \hline
  $B_1^{\pi}$ & $0.5$ & $-0.38(2)$ & $-0.56^{+0.06}_{-0.05}$ & $4454.3^{+2.7}_{-2.4}$ \\ 
  $B_1^{\pi}$ & $1.0$ & $-0.09(8)$ & $-1.10^{+0.21}_{-0.19}$ & $4451.5^{+2.6}_{-2.5}$ \\ 
  \hline \hline
\end{tabular}
\caption{
  The mass of the $P_c'$ pentaquark as deduced from the $E_a$ coupling extracted
  from the OBE model \mpv{in scenarios $A_1$ and $B_1$},
  both in the pionless and
  half-pionful theory.
  For comparison we remind that the location of the $\bar{D} \Lambda_{c1}$
  threshold in the isospin symmetric limit is $4459.5\,{\rm MeV}$.
  $E_a^{\rm crit}$ has the same meaning as in Table~\cite{tab:couplings-novel}.
}
\label{tab:M-Pc-prima-OBE}
\end{table}

\subsection{Can scenario $A$ be discarded?}

A preliminary examination of the different determinations of the couplings presented in Table~\ref{tab:couplings-novel} reveals that $D_b = 0$ in scenario $A$. 
The reason for this is that in general it is not possible to \mpv{{\it exactly}
  reproduce the masses of the two $\bar{D}^* \Sigma_c$ pentaquarks
  in this scenario}.
This seems counter intuitive at first, but actually there are good reasons
for this to be the case, which have to do with coupled-channel dynamics
and which we will explain below.
\mpv{Of course, we stress here that we are referring to the exact matching of
  the three known pentaquark masses with the three parameters of
  the present EFT: provided $D_b$ is smaller
  than its NDA estimate, the pentaquark trio is still
  well reproduced (particularly if we consider
  the experimental errors in the masses).}

First, we will consider a molecular pentaquark $P_Q$ in the heavy-quark limit, in which the masses of the charmed hadrons diverge and we can ignore the kinetic energy of the hadrons.
In this limit the binding energy of a molecular pentaquark is given by
\begin{eqnarray}
  B_{P_Q} = - \langle V^S \rangle \, ,
\end{eqnarray}
where $\langle V^S \rangle$ is the expected value of the S-wave potential and where we have taken the convention that the binding energy $B_{P_Q}$ is a positive number, thus the minus sign in front of $\langle V^S \rangle$.
Now we consider the case where the molecular pentaquark contains an additional P-wave component, for which the coupled-channel potential reads
\begin{eqnarray}
V_{P_Q} = 
\begin{pmatrix}
V^S & \lambda\,V^{SP} \\
\lambda\,V^{SP} & 0 
\end{pmatrix} \, ,
\end{eqnarray}
with $V^{SP}$ the S-to-P wave transition potential and $\lambda$ a number describing the strength of the transition potential.
If $\lambda$ is small, the effect of the coupled-channel dynamics on the binding of the pentaquarks can be estimated in perturbation theory, leading to
\begin{eqnarray}
  B_{P_Q} &=& - \langle V^S \rangle - \lambda^2 \, \langle V^{SP} G_0 V^{SP} \rangle 
  + \mathcal{O}(\lambda^3) \, ,
\end{eqnarray}
where $G_0$ is the two-hadron propagator, which in the static limit (infinitely heavy hadrons) reduces to
\begin{eqnarray}
  G_0 = \frac{1}{M_{P_Q} - M^P_{\rm th}} = \frac{1}{\Delta^P} \, ,
\end{eqnarray}
with $M_{P_Q}$ the mass of the heavy pentaquark, $M^P_{\rm th}$ the location of the P-wave threshold and $\Delta^P$ the energy gap.
This simplifies the S-to-P wave contribution to
\begin{eqnarray}
  B_{P_Q} &=& - \langle V^S \rangle - \lambda^2 \, \frac{\langle (V^{SP})^2 \rangle}{\Delta^P}  
  + \mathcal{O}(\lambda^3) \, ,
\end{eqnarray}
which will increase the binding energy provided that $\Delta^P < 0$, which happens to be the case~\footnote{Notice that we are assuming an attractive S-wave potential -- $\langle V^S \rangle < 0$ -- and that we always have $\langle (V^{SP})^2 \rangle \geqslant 0$.}.

The parameter $\lambda$ is useful because it is proportional to
the non-diagonal elements of the potentials
in Eqs.~(\ref{eq:V-Pc2}--\ref{eq:V-Pc-prima}).
Thus we have
\begin{eqnarray}
\lambda^2 \propto \mpv{\left \{ 6,4,1 \right \} }\, , \label{eq:lambda-2}
\end{eqnarray}
for \mpv{the $P_c'$, $P_c^*(1/2)$ and $P_c^*(3/2)$ pentaquarks}, respectively.
The actual effect of the P-wave channel also depends on the inverse of the mass gap, i.e.
\begin{eqnarray}
\Delta B_{P_Q} \propto -\frac{\lambda^2}{\Delta^P} \, ,
\label{eq:binding-gap}
\end{eqnarray}
which implies that the impact of the $\bar{D} \Lambda_{c1}$ channel will
be larger in the $P_c(4457)$ pentaquark than in the $P_c(4440)$ one
($\Delta_P = -2.2$ and $-19.2\,{\rm MeV}$ respectively).
However, once we take into account the finite mass of the hadrons, the effect
of the mass gap on the $P_c(4457)$ will diminish in relative terms
as it will be softened owing to the kinetic energy contributions.

\begin{figure}[ttt]
  \begin{center}
    \includegraphics[width=8.2cm]{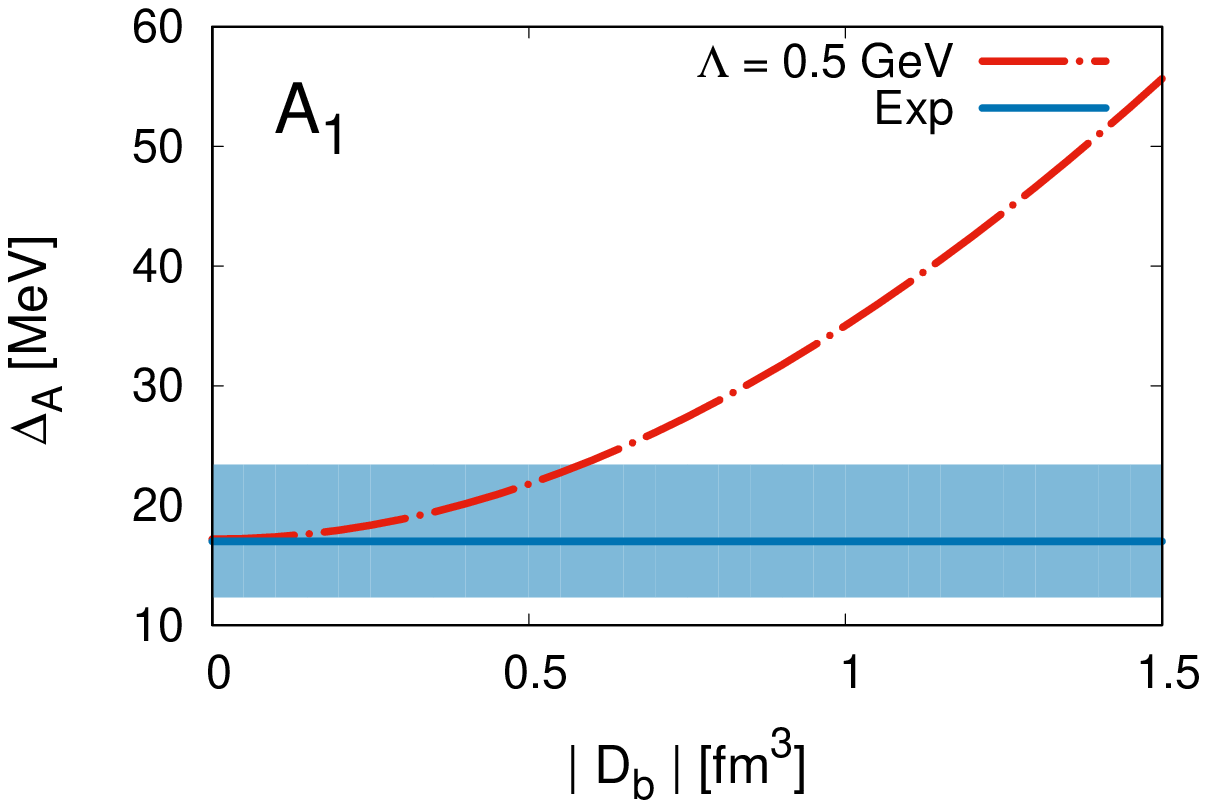}
    \includegraphics[width=8.2cm]{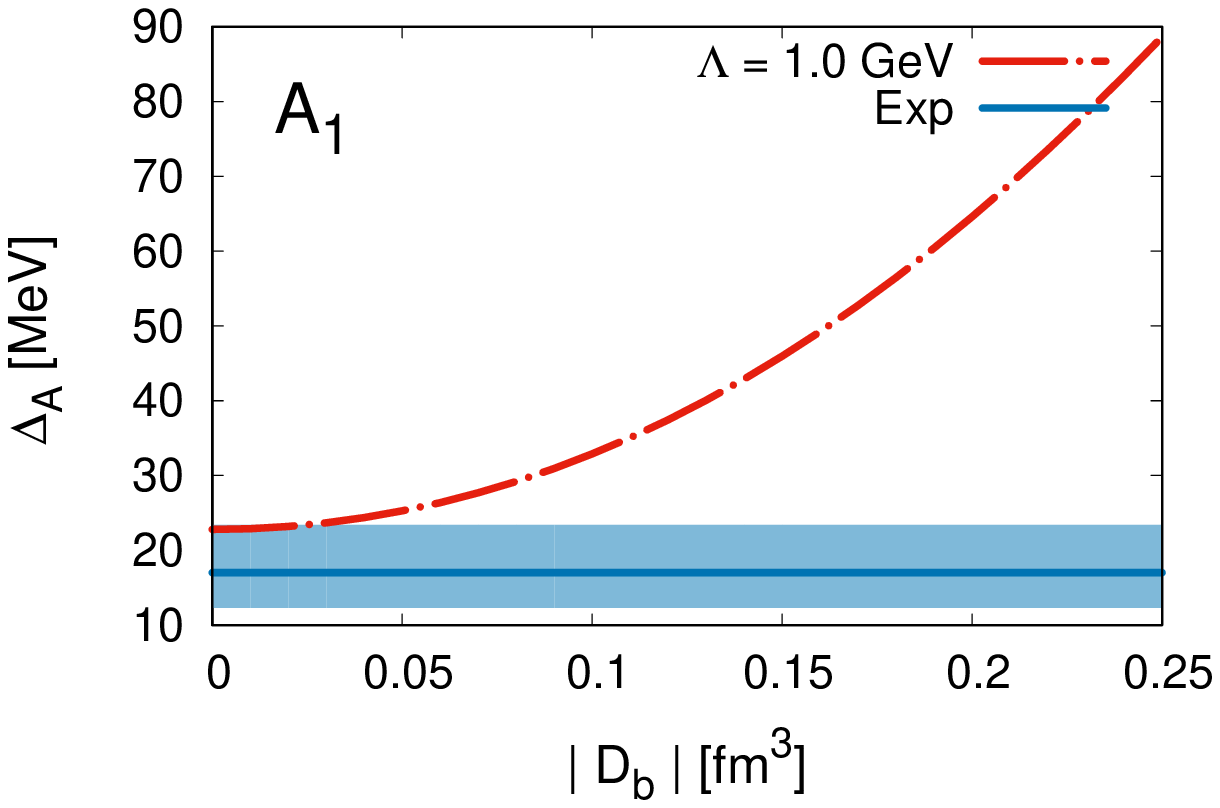}
\end{center}
\caption{
  Hyperfine splitting of the $P_c(4440)$ and $P_c(4457)$ pentaquarks
  in scenario $A$ as a function of $D_b$.
  We define the mass splitting as
  $\Delta = M(P_c^*(\frac{3}{2})) - M(P_c^*(\frac{3}{2}))$, the sign of which
  is positive (negative) in scenario $A$ ($B$).
  The blue band and blue line represent the experimental hyperfine splitting
  $| \Delta | = 17.0^{+6.4}_{-4.7}$.
  For simplicity we have done the calculation in the pionless EFT
  for $\Lambda =0.5$ and $1\,{\rm GeV}$, with the other
  EFT variations yielding similar results.
}
\label{fig:splitting-A}
\end{figure}

\begin{figure}[ttt]
  \begin{center}
    \includegraphics[width=8.2cm]{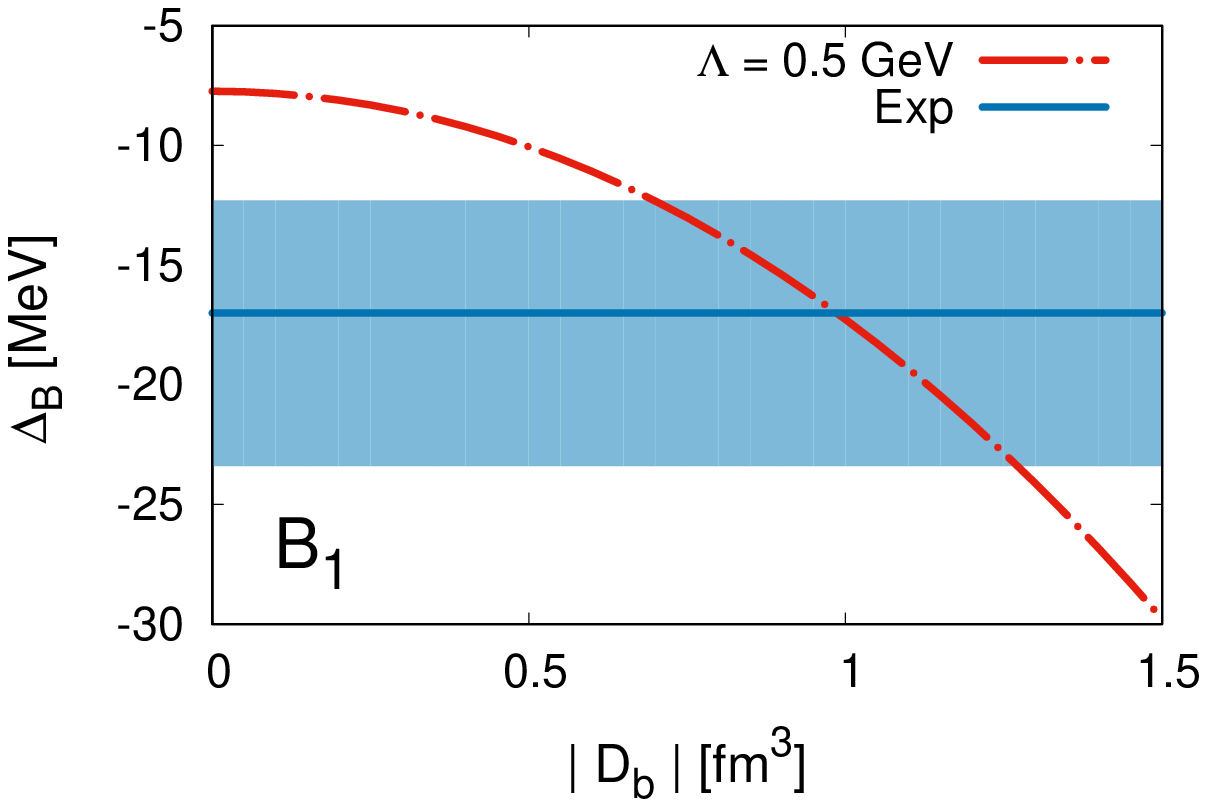}
    \includegraphics[width=8.2cm]{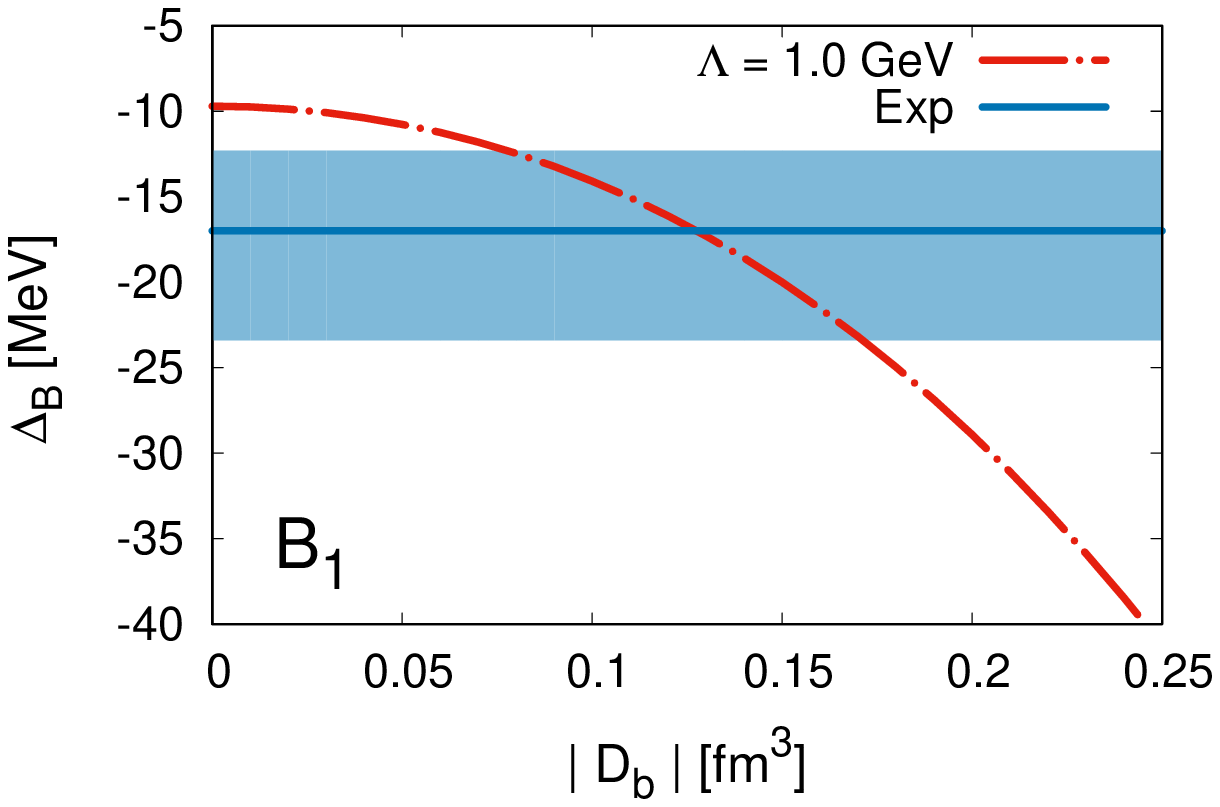}
\end{center}
\caption{
  Hyperfine splitting of the $P_c(4440)$ and $P_c(4457)$ pentaquarks
  in scenario $B$ as a function of $D_b$.
  The conventions are identical as in Fig.~\ref{fig:splitting-A}, to which
  we refer for further details.
}
\label{fig:splitting-B}
\end{figure}

In scenario $A$ \mpv{the $P_c(4440)$ and $P_c(4457)$ are already well reproduced
  before including the $\bar{D} \Lambda_{c1}$ channel,
  i.e. the choice $D_b = 0$ is compatible with the experimental
  location of the pentaquarks without further modifications.
  Thus it comes as no surprise that scenarios $A_0$ and $A_1$ are
  indistinguishable from the point of view of the couplings.
  Another factor to consider is the hyperfine splitting
  between the $P_c^*(1/2)$ and $P_c^*(3/2)$ pentaquarks,
  which in scenario $A_0$ is
  \begin{eqnarray}
    M_{A_0}(\frac{3}{2}) - M_{A_0}(\frac{1}{2}) = 17.2\,(22.8)\,{\rm MeV} \, ,
  \end{eqnarray}
  for the p-space Gaussian regulator with $\Lambda = 0.5\,(1.0)\,{\rm MeV}$,
  or
  \begin{eqnarray}
    M_{A_0}(\frac{3}{2}) - M_{A_0}(\frac{1}{2}) = 18.5\,(24.0)\,{\rm MeV} \, ,
  \end{eqnarray}
  for the r-space delta-shell regulator with $R_c = 1.0\,(0.5)\,{\rm fm}$.
  This is to be compared with the experimental splitting
  $|\Delta| = 17.0^{+6.4}_{-4.7}\,{\rm MeV}$, where we have combined
  the errors of the pentaquark masses in quadrature.
  Now it happens that the inclusion of the $\bar{D} \Lambda_{c1}$ channel
  widens the hyperfine splitting, as can be checked
  with the following calculation
  \begin{itemize}
  \item[(i)] determine the $C_a$ coupling from the location of
    the $P_c(4312)$ as a $\bar{D} \Sigma_c$ bound state (as usual),
  \item[(ii)] fix the coupling $D_b$ to a predetermined value,
  \item[(iii)] determine $C_b$ from the location $P_c(4457)$ as a
    $\bar{D}^* \Sigma_c$-$\bar{D} \Lambda_{c1}$ bound state and
  \item[(iv)] calculate the location of the $P_c(4440)$,
  \end{itemize}
  which is very similar to the procedures we have been following
  for scenarios $A_0$ and $B_0$.
  The splitting we obtain for scenario $A$ can be consulted
  in Fig.~\ref{fig:splitting-A}
  where for simplicity we have only shown the pionless calculation
  in p-space (the other calculations being qualitatively
  equivalent to this one).
  As can be appreciated in Fig.~\ref{fig:splitting-A}, the magnitude of
  the splitting grows with $D_b$, which in turn explains why it was not
  possible to find a solution in scenario $A_1$.
  However, once we consider the errors in the pentaquark masses,
  scenario $A$ is still compatible with the experimental location of
  the pentaquarks for small values of the coupling $D_b$.
  For the case of $\Lambda = 0.5\,(1.0)\,{\rm GeV}$, scenario $A$ is compatible
  with experiment provided $|D_b| < 0.58 \,(0.02)\,{\rm fm}^3$, which is
  to be compared with the NDA estimate $|D_b^{\rm NDA}| = 0.10\,{\rm fm}^3$.
  This indicates that NDA and scenario $A$ are compatible
  for $\Lambda = 0.5\,{\rm GeV}$, but not for $\Lambda = 1.0\,{\rm GeV}$.

  For scenario $B$ the hyperfine splitting is
  \begin{eqnarray}
    M_{B_0}({\frac{3}{2}}) - M_{B_0}({\frac{1}{2}}) = -7.7\,(-9.7)\,{\rm MeV} \, ,
  \end{eqnarray}
  for the p-space Gaussian regulator with $\Lambda = 0.5\,(1.0)\,{\rm MeV}$,
  or
  \begin{eqnarray}
    M_{B_0}({\frac{3}{2}}) - M_{B_0}({\frac{1}{2}}) =
    -8.1\,(-10.4)\,{\rm MeV} \, ,
  \end{eqnarray}
  for the r-space delta-shell regulator with $R_c = 1.0\,(0.5)\,{\rm fm}$,
  to be compared with the $-17.0\,{\rm MeV}$ figure which we obtain
  from the experimental masses.
  The inclusion of the $\bar{D} \Lambda_{c1}$ channel actually increases
  the absolute magnitude of the mass splitting, thus improving
  the agreement between theory and experiment.
  This can be seen in Fig.~\ref{fig:splitting-B}, where we show the hyperfine
  splitting as a function of the coupling $D_b$ for scenario $B$.
  The details of the calculation are analogous to the ones we followed
  for Fig.~\ref{fig:splitting-A}.
  Compatibility with the size of the experimental splitting is possible
  for $0.70\,{\rm fm}^3 < | D_b | < 1.26\,{\rm fm}^3$
  ($0.08\,{\rm fm}^3 < | D_b | < 0.17\,{\rm fm}^3$)
  for $\Lambda = 0.5\,(1.0)\,{\rm GeV}$.
  This indicates that scenario $B$ agrees with the NDA estimates of $D_b$
  for $\Lambda = 1.0\,{\rm GeV}$, while for $\Lambda = 0.5\,{\rm GeV}$
  it requires a $D_b$ seven times the NDA estimate.

  At this point we observe that the NDA estimates for $D_b$ seem to be better
  respected for $\Lambda = 1.0\,{\rm GeV}$ than
  for $\Lambda = 0.5\,{\rm GeV}$.
  We are not only referring to Figs.~\ref{fig:splitting-A}
  and ~\ref{fig:splitting-B}, but also to the couplings
  in Table \ref{tab:couplings-novel}.
  Actually this might be related to a series of factors, the most important
  one probably being that naturalness of contact-range interactions
  beyond S-waves has been rarely dealt with in the literature,
  except in Halo EFT~\cite{Bertulani:2002sz} as previously noted.
  However there are at least two reasons to revise the NDA estimates
  for $D_b$ upwards.
  The first one has to do with the numerical factors in front of $D_b$.
  The NDA estimate for $C_{l,l'}$ implicitly assumes a contact-range potential
  normalized as
  \begin{eqnarray}
    \langle p'(l') | V_C | p(l) \rangle = C_{l,l'} {p'}^{l'} {p}^l \, .
  \end{eqnarray}
  In contrast, the $\bar{D}^* \Sigma_c \to \bar{D} \Sigma_c$ transition
  potentials contain a factor in front of the coupling
  \begin{eqnarray}
    \langle p'(0) | V_C | p(1) \rangle =\lambda\,
    \frac{2}{3 \sqrt{3}}\,D_b\,{p} \, ,
  \end{eqnarray}
  with $\lambda$ a numerical factors that can be deduced
  from Eq.~(\ref{eq:lambda-2}).
  Combining all the factors, this indicates that the NDA value of $D_b$
  should be $1.1-2.6$ times the value of
  $C^{\rm NDA}_{\rm SP} = 0.10\,{\rm fm}^3$,
  depending on which pentaquark we use as a reference.
  The second reason has to do with the choice of a hard scale $M$, for which
  we previously took the $1\,{\rm GeV}$ figure.
  Had we taken $M = m_{\rho}$ with $m_{\rho} = 770\,{\rm MeV}$ the rho meson mass,
  we would have arrived at $C^{\rm NDA}_{\rm SP} = 0.21\,{\rm fm}^3$,
  i.e. twice the estimation for $M = 1\,{\rm GeV}$.
  Putting all these pieces together would suggest
  $D_b^{\rm NDA} = (0.10-0.55)\,{\rm fm}^3$.
  But there could also be unaccounted reasons for $D_b$ to be smaller.
  The bottom-line is that we still require external input to decide which
  scenario is more probable, be it phenomenological studies or
  further experiments.
}

Here it is important to mention that the two theoretical scenarios
we have presented ($A$ and $B$) are but a subset of all the possible scenarios.
We have three molecular explanations ($P_c^*(1/2)$, $P_c^*(3/2)$ and $P_c'$) for two pentaquarks,
which gives a total of six possible scenarios
instead of the two we are considering.
But with the exception of scenarios $A$ and $B$, it is not possible
to determine the value of the couplings in other cases.
For instance, had we assumed that the $P_c(4440)$ is the $J=\tfrac{3}{2}$
$\bar{D}^* \Sigma_c$ molecule and $P_c(4457)$ the $J=\tfrac{1}{2}$
$\bar{D} \Lambda_{c1}$ one, i.e. the scenario originally proposed in Ref.~\cite{Burns:2019iih}, we would have ended with three unknown
couplings ($C_b$, $D_b$ and $E_a$) for two pentaquarks.
Though this limitation \mpv{might} indeed be overcome by invoking NDA,
the resulting analysis is considerably more involved
than in scenarios $A$ and $B$ and thus we have decided
not to consider them in this work.

Another factor that we have not taken into account in the present analysis
is the effect of the $\bar{D} \Lambda_{c1}^*$ channel,
which lies about $20\,{\rm MeV}$ above the $\bar{D} \Lambda_{c1}$ threshold.
The $\bar{D} \Lambda_{c1}^*$ channel can mix with the
$J=\tfrac{3}{2}$ $\bar{D}^* \Sigma_c$ one,
inducing a bit of extra attraction
in this later case.
However, from Eq.~(\ref{eq:binding-gap}) and the larger mass gap for the $\bar{D} \Lambda_{c1}^*$ channel ($\Delta_P = -55.0\,{\rm MeV}$ versus $-19.2\,{\rm MeV}$ for the $\bar{D} \Lambda_{c1}$ one for scenario $B$), we expect this effect to be fairly modest.

\section{Conclusions}
\label{sec:conclusions}

In this manuscript we have considered the impact of
the $\bar{D} \Lambda_{c1}$ channel for the description of
the $P_c(4440)$ and $P_c(4457)$ pentaquarks.
Within the molecular picture, the standard interpretation of
the $P_c(4440)$ and $P_c(4457)$ states is that
they are $\bar{D}^* \Sigma_c$ bound states.
This is motivated by the closeness of the $\bar{D}^* \Sigma_c$ threshold
to the location of the two pentaquark states.
But the same is true for the $\bar{D} \Lambda_{c1}$ threshold, which
naturally prompts the question of what is the contribution of
this channel to the description of
the pentaquarks~\cite{Burns:2015dwa,Geng:2017hxc,Burns:2019iih}.

For answering this question we have analyzed the inclusion of
$\bar{D} \Lambda_{c1}$ channel from the EFT perspective.
We find that the importance of the $\bar{D} \Lambda_{c1}$ channel
depends on which are the quantum numbers of
the $P_c(4440)$ and $P_c(4457)$ pentaquarks:
in the standard molecular interpretation ($\bar{D}^* \Sigma_c$) their quantum
numbers can be either $J^P = \tfrac{1}{2}^-$ or $\tfrac{3}{2}^-$,
but we do not know which quantum numbers correspond to which pentaquark.
There are two possibilities: that the $P_c(4440)$ and $P_c(4457)$ are
respectively the $J^P = \tfrac{1}{2}^-$ and
$\tfrac{3}{2}^-$ $\bar{D}^* \Sigma_c$ molecules, or vice versa.
The first possibility, which we call scenario $A$, corresponds to
the standard expectation that hadron masses increase with spin.
The second possibility, scenario $B$, represents the opposite pattern,
which has recently been conjectured to be a property of
hadronic molecules~\cite{Peng:2020xrf}.

In scenario $A$ the inclusion of the $\bar{D} \Lambda_{c1}$ channel is
inconsequential for the description of the molecular pentaquarks:
the $\bar{D} \Lambda_{c1}$ can effectively be ignored,
as the transition potential between the
$\bar{D} \Lambda_{c1} \to \bar{D}^* \Sigma_c$ channels
\mpv{is required to be weak if we want to reproduce
  the three known pentaquarks.}
However this is not the case in scenario $B$, where the inclusion of
the $\bar{D} \Lambda_{c1}$ channel can potentially have important
consequences on the pentaquark spectrum.
In this case the coupling between the $\bar{D} \Lambda_{c1}$ and
$\bar{D}^* \Sigma_c$ channels \mpv{might very well be} strong enough
as to facilitate the binding of the $\bar{D} \Lambda_{c1}$ system
in S-wave, as happened in Ref.~\cite{Burns:2019iih}.
Right now there is no experimental determination of 
the quantum numbers of the pentaquarks, with different theoretical
explorations favoring different scenarios. We see a preference
towards A in Refs.~\cite{Liu:2019tjn,Xiao:2019aya} and
towards B in Refs.~\cite{Shimizu:2019ptd,Yamaguchi:2019seo,Valderrama:2019chc,Du:2019pij},
though other scenarios are possible: for instance in Ref.~\cite{Burns:2019iih} the $\tfrac{1}{2}^-$ $\bar{D}^* \Sigma_c$ pentaquark does not bind.
Within the molecular picture there seems to be a tendency
for pionless theories to favor $A$,
while theories that include pion exchange effects tend to fall
into scenario $B$.

If scenario $B$ happens to be the one preferred by nature,
the prospects for the $J^P = \tfrac{1}{2}^+$ $\bar{D} \Lambda_{c1}$ molecule
to bind are good: though the fate of this bound state is ultimately
contingent on the unknown short-distance details of the interaction,
phenomenological arguments indicate a moderate attraction
between the $\bar{D}$ meson and $\Lambda_{c1}$ baryon
at short distances.
If this is the case and this molecule binds, it might very well be that
the $P_c(4457)$ is a double peak, containing both a $\bar{D}^* \Sigma_c$
and a $\bar{D} \Lambda_{c1}$ molecule with opposite parities.
If scenario $A$ is the one that actually describes the pentaquarks,
the $J^P = \tfrac{1}{2}^+$ pentaquark cannot be discarded either
--- there is the possibility that it binds even without coupling to
the $\bar{D}^* \Sigma_c$ channel --- but is less likely
to exist nonetheless.

Yet we stress the exploratory nature of the present manuscript:
the EFT framework requires experimental input and a series of
assumptions for it to be able to generate predictions.
\mpv{Indeed there could be three pentaquarks independently of the quantum
  numbers of the experimentally observed ones, with scenario $B$ merely
  being the case for which this possibility is more probable.}
In this regard it would be very welcome to have phenomenological explorations
of the $\bar{D} \Lambda_{c1}$ interaction and
the $\bar{D} \Lambda_{c1} \to \bar{D}^* \Sigma_c$ transition.

\section*{Acknowledgments}

This work is partly supported by the National Natural Science Foundation
of China under Grants No. 11735003, 
11975041, the Fundamental Research Funds for the Central Universities
and the Thousand Talent Plan for Young Professionals.

\appendix
\section{Pion- and Rho-like couplings}
\label{app:saturation}

In this appendix we discuss the possible sources of saturation of
the $D_a$ and $D_b$ contact-range couplings that mediate
the $\bar{D}^* \Sigma_c \to \bar{D} \Lambda_{c1}$
transition.

Regarding the coupling $D_a$, its similarity with the exchange of a pseudoscalar
is evident from a direct comparison to the contact-range
potential of Eq.~(\ref{eq:V-C2-full}) for $D_b = 0$, that is:
\begin{eqnarray}
  V_{C2(a)}(1 \to 2) = + D_a \, \vec{\sigma}_{L1} \cdot \vec{q} \, ,
\end{eqnarray}
and the OPE potential of Eq.~(\ref{eq:OPE}).
Saturation of the $D_a$ coupling from a derivative pseudoscalar meson
such as the pion will lead to the approximation~\cite{Lu:2017dvm,Peng:2020xrf}
\begin{eqnarray}
  D_a^{(\pi)} \propto
  \frac{g_1 h_2}{\sqrt{2} f_{\pi}^2}
  \,\vec{\tau}_1 \cdot \vec{t}_2\,
  \frac{\omega_{\pi}}{\mu_{\pi}^2} \, .
\end{eqnarray}
However saturation is only known to work if the regularization scale
is close to the mass of the exchanged meson~\cite{Epelbaum:2001fm}.
Taking into account that the pion is the lightest meson and that the cutoff
range we are using is $\Lambda = (0.5-1.0) \,{\rm GeV}$,
we do not expect the $D_a$ coupling to receive
contributions coming from pion exchange.
If we consider the exchange of heavier mesons, there is no clear candidate
for the exchange of a pseudoscalar meson in the mass range comprised
by our choice of a cutoff.
Hence the decision to consider that $D_a = 0$ at lowest order.

For the $D_b$ coupling the situation is different, because
the $\bar{D}^* \Sigma_c \to \bar{D} \Lambda_{c1}$ transition
can happen via rho-exchange.
The relevant Lagrangians read
\begin{eqnarray}
  \mathcal{L}_{\rho 1} &=& 
          {g}_{\rho 1}\,q_{L}^{\dagger} \, {\tau}_a \, {\rho}_{a 0} \, q_{L}
          \nonumber \\
          &-& \frac{f_{\rho 1}}{4 {M}} \epsilon_{ijk}
          q_{L}^{\dagger}\, \sigma_{L, k} {\tau}_a \cdot
          ( \partial_i {\rho}_{a j} - \partial_j {\rho}_{a i} )\, q_{L} \, , \\
  \mathcal{L}_{\rho 2} &=& \frac{f_{\rho 2}}{2 M}\,
  {v}_L^{\dagger} \,t_a ({J}_{L i} \partial_i \rho^0_a  - \partial_0 {\rho}_{a i})\, {a}_L + C.C. \, 
\end{eqnarray}
where $q_L$, $a_L$ and $v_L$ are the light subfields of
the $D^{(*)}$, $\Sigma_c^{(*)}$ and $\Lambda_{c1}^{(*)}$ charmed hadrons,
$\rho_{a \mu}$ is the rho meson field, with $\mu$ a Lorentz index
($i$ is used to indicate $\mu = 1,2,3$) and $a$ and isospin index,
$t_a$ and $\tau_a$ are isospin matrices, $g_{\rho 1}$, $f_{\rho 1}$ and $f_{\rho 2}$
are coupling constant, and $M$ is a mass scale for the magnetic
and electric dipole terms (i.e. the piece proportional to
$f_{\rho 1}$ and $f_{\rho 2}$, respectively).
The charge-like term (i.e. the one proportional to $g_{\rho 1}$)
can contribute to $\bar{D} \Sigma_c \to \bar{D} \Lambda_{c1}$ and
$\bar{D}^* \Sigma_c \to \bar{D}^* \Lambda_{c1}$ transitions,
but not to the $\bar{D}^* \Sigma_c \to \bar{D} \Lambda_{c1}$
one which is of interest for this work.
The magnetic and electric and dipole terms of these Lagrangian
lead to the potential
\begin{eqnarray}
  V_{\rho}(\vec{q}, 1 \to 2) &=& \frac{f_{\rho 1}}{2 M}\,\frac{f_{\rho 2}}{2 M}\,
  \vec{\tau}_1 \cdot \vec{t}_2\,
  \frac{\omega_{\rho}}{{\vec{q}\,}^2 + \mu_{\rho}^2}\,
  \vec{q} \cdot \left( \vec{\sigma}_{L1} \times \vec{J}_{L2}\right) \, ,
  \nonumber \\
\end{eqnarray}
where $\omega_{\rho} \simeq (m({\Lambda_{c1}}) - m(\Sigma_c)) \simeq (m(D^*) - m(D))$,
$\mu_{\rho}^2 = m_{\rho}^2 - \omega_{\rho}^2$ and the rest of the terms
have the same meaning as in Eqs.~(\ref{eq:V-C2-full}) and (\ref{eq:OPE}).
Finally the saturation of the $D_b$ coupling by the rho will lead to a
value proportional to
\begin{eqnarray}
  D_b^{(\rho)} \propto
  \frac{f_{\rho 1}}{2 M}\,\frac{f_{\rho 2}}{2 M}\,
  \vec{\tau}_1 \cdot \vec{t}_2\,
  \frac{\omega_{\rho}}{\mu_{\rho}^2} \, .
\end{eqnarray}
This is why we keep $D_b$ as a leading-order effect, but consider $D_a$
to be subleading.


%

\end{document}